\documentclass[sigconf,screen,nonacm]{acmart}
\AtBeginDocument{
  }
\usepackage{enumitem}
\usepackage[utf8]{inputenc}
\usepackage{booktabs}
\usepackage{multirow}
\usepackage{makecell}

\usepackage{amssymb}
\usepackage{newtxmath}
\usepackage[table]{xcolor}
\usepackage{tabularray}
\usepackage{placeins}
\usepackage{caption}
\usepackage{algorithm}
\usepackage{algorithmic}
\usepackage{bbm}
\usepackage{float}

\setcopyright{none}
\begin{document}

\title[MMDynOpt-Agent: Dynamic Optimization Framework for Multimodal Large Language Model Reasoning]{MMDynOpt-Agent: Dynamic Optimization for Multimodal Large Language Model Reasoning via Reinforcement Learning}

\author{
  \bfseries\large
  Wenjin Liu\textsuperscript{1,2},
  Haoran Luo\textsuperscript{1,$\dagger$},
  Fayuan Ke\textsuperscript{2},
  Zhenghong Lin\textsuperscript{1},
  Yue Lu\textsuperscript{2},\\
  Zhe Cui\textsuperscript{2,$\dagger$},
  Anh Tuan Luu\textsuperscript{1},
  Carl Yang\textsuperscript{3}
}
\affiliation{
  \institution{
    \large
    \textsuperscript{1}Nanyang Technological University, College of Computing and Data Science, Singapore\\
    \textsuperscript{2}Hithink Research, Code Generation Department, China\\
    \textsuperscript{3}Emory University, Department of Computer Science, United States\\
    {\normalsize\texttt{wenjinliu23@outlook.com, haoran.luo@ntu.edu.sg, kefayuan123@gmail.com,}}\\
    {\normalsize\texttt{hongzhenglin970323@gmail.com, luyue3@myhexin.com, cuizhe@myhexin.com,}}\\
    {\normalsize\texttt{anhtuan.luu@ntu.edu.sg, j.carlyang@emory.edu}}
  }
  \country{}
}

\renewcommand{\shortauthors}{Wenjin Liu et al.}

\begin{abstract}
Recently, multimodal large language models (MLLMs) have demonstrated strong potential in visual understanding and complex reasoning tasks. However, existing methods often struggle to efficiently transform visual cues from multimodal inputs and the semantics of the question into effective reasoning conditions, thereby limiting the reasoning performance of multimodal large language models. To address this challenge, we propose MMDynOpt-Agent, which models the dynamic optimization of multimodal reasoning as a Markov decision process via end-to-end reinforcement learning. Specifically, a lightweight multimodal agent serves as the decision policy and interacts with the target MLLM as the environment, adaptively steering its reasoning through multi-turn dynamic optimization prompts. Furthermore, to reduce the cost of multimodal reasoning, a reward mechanism that combines format compliance, answer correctness, and budget awareness is designed to jointly ensure reasoning accuracy and efficiency. MMDynOpt-Agent is transferable and generalizable, enabling training with one target MLLM and inference-time transfer to others. Experimental results on fifteen public datasets show MMDynOpt-Agent achieves strong performance and outperforms baselines.
Our project is available\footnote{\url{https://github.com/QwenQKing/MMDynOpt-Agent}}.
\end{abstract}

\ccsdesc[500]{Computing methodologies~Knowledge representation and reasoning}
\ccsdesc[500]{Theory of computation~Semantics and reasoning}
\ccsdesc[500]{Computing methodologies~Artificial intelligence}

\keywords{Multimodal Large Language Model, Dynamic Optimization, Multimodal Reasoning, Efficient Reasoning, Reinforcement Learning, Markov Decision Process}

\maketitle
\begingroup
\renewcommand{\thefootnote}{}
\footnotetext{\textsuperscript{$\dagger$}Corresponding authors.}
\endgroup
\section{Introduction}
\label{introduction}
In recent years, MLLMs \cite{achiam2023gpt,bai2025qwen3,wu2024deepseek} have shown strong potential in visual question answering \cite{huang2025visual,wang2024charxiv}, financial computation \cite{li2025fundamental}, medical diagnosis \cite{wang2025survey}, and scientific reasoning \cite{sun2025survey}. By combining visual encoding with the knowledge and reasoning abilities of large language models (LLMs), they support cross-modal integration and multi-step reasoning \cite{caffagni2024revolution,liu2025prompt}. However, unlike traditional visual understanding, multimodal reasoning requires not only perception but also semantically structured visual evidence and intermediate reasoning conditions \cite{jiang2025specific,doddapaneni2025primer}. Therefore, improving the ability of MLLMs to leverage visual information and semantic constraints for accurate reasoning in complex tasks remains challenging \cite{yang2026model}.

\begin{figure}[t]
  \centering
  \includegraphics[width=1\linewidth]{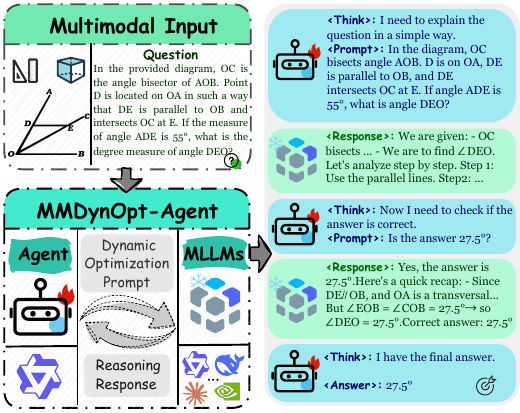}
  \caption{An example of MMDynOpt-Agent steering the reasoning of the target MLLM through multi-turn dynamic optimization prompts, refining and verifying the answer.}
    \label{fig: agent_case}
\end{figure}

To enhance the reasoning of MLLMs, methods based on post-training \cite{zhang2026instruction}, prompt optimization \cite{trivedi2025align}, test-time scaling (TTS) \cite{zhang2025survey}, and multi-agent systems (MAS) \cite{he2025llm} have been proposed. Post-training includes supervised fine-tuning (SFT) and reinforcement learning (e.g., Group Relative Policy Optimization \cite{guo2025deepseek}) for instruction alignment and reasoning. Prompt optimization comprises structured prompting (e.g., Chain-of-Thought \cite{wei2022chain}) via templates and automatic prompt optimization (e.g., Genetic-Pareto \cite{agrawal2025gepa}) via search or evolutionary algorithms. TTS spans sampling-and-aggregation (e.g., Best-of-N \cite{brown2024large}), tree-based search (e.g., Tree of Thoughts \cite{yao2023tree}), and iterative revision (e.g., Self-Refine \cite{madaan2023self}), improving reliability through candidate expansion, step-level exploration, and feedback. MAS include division-of-labor collaboration (e.g., MetaGPT \cite{hong2023metagpt}) and debate-and-critique interaction (e.g., Multi-agent Debate \cite{du2024improving}), emphasizing information aggregation among agents \cite{torreno2017cooperative}.

\begin{figure*}[t]
  \centering
  \includegraphics[width=1\linewidth]{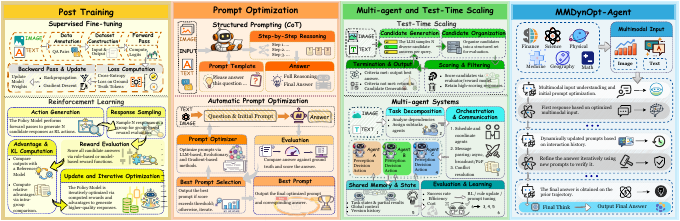}
  \caption{Comparison of different methods for improving the reasoning performance of MLLMs: post training,  prompt optimization, multi-agent and test-time scaling, as well as our dynamic prompt optimization framework, MMDynOpt-Agent.}
 \label{fig: four_methods}
\end{figure*}

However, these methods still face three challenges:
\textbf{(i) Lack of dynamic adaptability to multimodal inputs.} Visual information is often sparse and heterogeneously distributed, and static prompts cannot dynamically adjust reasoning conditions for both image and question inputs, causing MLLMs to overlook critical visual regions and semantic cues during multi-step reasoning.
\textbf{(ii) Difficulty in balancing reasoning quality and efficiency.} Although multi-turn and collaborative reasoning can improve reasoning quality, they also increase interaction rounds and token consumption, making it difficult to balance reasoning quality with efficiency.
\textbf{(iii) Limited transferability and generalization of optimization strategies.} Existing optimization strategies depend on model-specific representations and reasoning behaviors, lacking standardized interfaces for dynamic optimization, which constrains their transfer to other MLLMs and cross-model generalization.

To address these challenges, we propose MMDynOpt-Agent (see Figure \ref{fig: agent_case}), an end-to-end reinforcement learning-based framework for MLLM reasoning optimization. MMDynOpt-Agent generates dynamic prompts by analyzing the image and question, transforming multimodal inputs into reasoning conditions for external large-scale models, enhancing reasoning accuracy and information use across rounds. Furthermore, we design Prism Reward, which jointly constrains format compliance, answer correctness, and budget awareness, allowing the agent to adaptively adjust rounds and token usage while ensuring quality, thus achieving an efficient performance-resource balance. In addition, the agent can be applied to diverse large-scale MLLMs, demonstrating improved cross-model reasoning and generalization, and establishing a dynamic, cost-aware, transferable optimization mechanism for complex tasks.

We evaluate MMDynOpt-Agent on fifteen datasets covering multimodal reasoning tasks in financial, medical, and scientific domains. Results indicate MMDynOpt-Agent outperforms baselines (see Figure \ref{fig: four_methods}) in answer quality and reasoning performance. The trained agent transfers to MLLMs, maintaining performance on unseen data and models, suggesting robust generalization. Overall, MMDynOpt-Agent provides an efficient and reliable solution for multimodal reasoning tasks, demonstrating potential for application.

\section{Related Work}
\label{related_work}

\textbf{Post Training.} 
Post training includes parameter-efficient fine-tuning (PEFT) and reinforcement learning (RL). PEFT reduces trainable parameters via low-rank decomposition (Low-Rank Adaptation \cite{hu2022lora}, Quantized Low-Rank Adaptation \cite{dettmers2023qlora}), bottleneck modules (Adapter \cite{houlsby2019parameter}), and soft prompts \cite{li2021prefix,ding2022delta}. RL optimizes reasoning via reward signals, evolving from Reinforcement Learning from Human Feedback \cite{christiano2017deep} and Proximal Policy Optimization (PPO) \cite{schulman2017proximal} to Direct Preference Optimization (DPO) \cite{rafailov2023direct} and Group Relative Policy Optimization (GRPO) \cite{guo2025deepseek}, reducing training overhead.

\textbf{Prompt Optimization.} 
Prompt optimization includes structured prompting and automatic prompt optimization (APO) \cite{du2026survey,ramnath2025systematic,liu2026systematic,zhang2026instruction}. Structured prompting organizes reasoning through templates, including Chain-of-Thought (CoT) \cite{wei2022chain}, Least-to-Most prompting \cite{zhou2022least}, and multimodal Compositional CoT \cite{mitra2024compositional}. APO includes Optimization by PROmpting (OPRO) \cite{yang2023large} and PromptAgent \cite{wang2023promptagent} for instruction discovery, while TextGrad \cite{yuksekgonul2024textgrad} and Genetic-Pareto (GEPA) \cite{agrawal2025gepa} introduce textual gradients and genetic reflection.

\textbf{Test-Time Scaling.}
TTS enhances MLLMs by adding inference-time computation without parameter updates \cite{snell2025scaling,liu2025rethinking}. Sampling-aggregation expands candidates and aggregates outputs via majority voting, including Self-Consistency \cite{wang2022self}, Best-of-N \cite{brown2024large}, and $\phi$-Decoding \cite{xu2025phi}. Reward-guided search uses process reward models (PRMs): Let's Verify Step by Step \cite{lightman2023let} and Math-Shepherd \cite{wang2024math} establish human- and auto-annotated PRMs, while ToT \cite{yao2023tree}, ReST-MCTS* \cite{zhang2024rest}, and AFLOW \cite{zhang2024aflow} combine tree search, PRM-guided self-training, and workflow discovery. Self-Refine \cite{madaan2023self} enables training-free refinement via generation–revision cycles.

\begin{figure*}[t]
  \centering
  \includegraphics[width=1\linewidth]{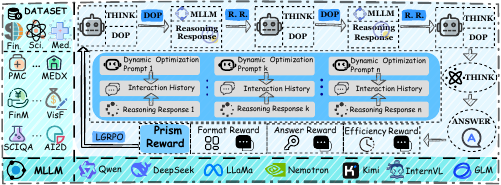}
  \vspace{-5mm}
  \caption{Overview of the MMDynOpt-Agent framework. In this framework, a multimodal agent is trained via reinforcement learning to dynamically generate optimized prompts for MLLMs. DOP: dynamic optimization prompt. R. R.: reasoning response. }
    \label{fig: MMDynOpt-Agent}
\end{figure*}

\textbf{Multi-agent Systems.}
MAS enhance MLLM reasoning through collaboration and critique \cite{gao2024multi,chen2025confluence}. Collaboration includes SOP-driven pipelines (MetaGPT \cite{hong2023metagpt}), dynamic team formation (AgentVerse \cite{chen2023agentverse}), shared reflection (COPPER \cite{bo2024reflective}), and adaptive allocation (MDAgents \cite{kim2024mdagents}); critique includes debate and voting (Multiagent Debate \cite{du2024improving}, MAD \cite{liang2024encouraging}, ChatEval \cite{chan2023chateval}, ReConcile \cite{chen2024reconcile}). Recently, layered aggregation (MoA \cite{wang2024mixture}), cooperative reinforcement learning (CORY \cite{ma2024coevolving}), Socratic prompt optimization (MARS \cite{zhang2025mars}), and MacNet \cite{qian2024scaling} show logarithmic-style gains with agent scaling.

\section{Problem Formulation}

MMDynOpt-Agent formulates multimodal collaborative reasoning as a dynamic optimization process in which the agent guides the large-scale MLLM and optimizes the steering policy via end-to-end reinforcement learning. This process is formalized as follows.

\textbf{Task Definition.}
Given a multimodal input $q = (q_{\text{text}}, q_{\text{img}})$, where $q_{\text{text}}$ is the question and $q_{\text{img}}$ is the image, the goal is to obtain the answer $\hat{y}$. The system consists of two core components:

\textit{(i) Multimodal Agent} $\pi_\theta$: a lightweight MLLM as the agent that analyzes the multimodal input $q$ and generates steering prompts to dynamically optimize the reasoning input of the large-scale MLLM.

\textit{(ii) Target MLLM} $\mathcal{E}$: a target MLLM acts as the environment that receives the agent's instructions and returns reasoning results.

\textbf{Dynamic Optimization Paradigm.}
The agent $\pi_\theta$ dynamically optimizes the reasoning of MLLM $\mathcal{E}$ through up to $T$ dynamic optimization rounds. At each round, $\pi_\theta$ generates an action $a_t$: a \emph{steering action} sends a dynamic optimization prompt to $\mathcal{E}$ to refine reasoning, while an \emph{answering action} produces the answer. Upon a steering action, $\mathcal{E}$ returns an observation $o_t$ that is incorporated into the next-round state, yielding a dynamic optimization trajectory:
\begin{equation}
    \tau = (s_0, a_0, o_0, s_1, a_1, o_1, \ldots, s_{T_\tau}, a_{T_\tau})
\end{equation}
where $s_t$ is the state at round $t$, and $T_\tau \leq T$ is the number of rounds.

\textbf{Optimization Objective.}
The agent policy $\pi_\theta$ is optimized via reinforcement learning to maximize the trajectory-level reward:
\begin{equation}
    \pi_\theta^* = \arg\max_{\pi_\theta}\ \mathbb{E}_{q \sim \mathcal{D}}\left[R(\tau \mid \pi_\theta, \mathcal{E}, q)\right]
\end{equation}
where $\mathcal{D}$ is the training data distribution, and $R(\tau)$ jointly measures format compliance, answer correctness, and reasoning efficiency.

\section{Methodology}
\label{methodology}
In this section, we introduce the details of the MMDynOpt-Agent (see Figure \ref{fig: MMDynOpt-Agent}), including Markov decision-driven dynamic optimization, Prism reward-guided policy optimization, and the transferable dynamic optimization framework for collaborative reasoning.

\definecolor{tagthink}{rgb}{0.1,0.3,0.9}
\definecolor{tagprompt}{rgb}{0.0,0.7,0.8}
\definecolor{taganswer}{rgb}{0.5,0.0,0.8}
\definecolor{question}{rgb}{0.9,0.1,0.1}
\begin{table*}[t]
\centering
\caption{The initial prompt template utilized by the agent $\pi_\theta$ to dynamically optimize the prompts provided to the MLLM $\mathcal{E}$.}
\label{tab:sys-prompt-tem}
\vspace{-0.8em} 
\begin{minipage}{\textwidth}
\hrule height 1pt
\vspace{0.36em}
\fontsize{8.46}{8}\selectfont
First, provide a simple explanation for the question to another more powerful large language model for a correct answer. Focus on explaining the question without deep reasoning in the first step. After receiving the response from the large language model, think about the response, and by interacting with the large language model again and again, arrive at the final answer. Proceed step by step with the following rules:
\textbf{1.} Only in the first interaction, provide a brief explanation of the question and give it to the large language model:
\texttt{\textbf{\textcolor{tagthink}{<think>}}} (briefly think without deep reasoning) \texttt{\textbf{\textcolor{tagthink}{</think>}}}.
\texttt{\textbf{\textcolor{tagprompt}{<prompt>}}} (give the question and its explanation to the large language model) \texttt{\textbf{\textcolor{tagprompt}{</prompt>}}}.
\textbf{2.} After the first interaction, in each interaction with the large language model, you must:
\texttt{\textbf{\textcolor{tagthink}{<think>}}} (reason about the current response and the question) \texttt{\textbf{\textcolor{tagthink}{</think>}}}.
\texttt{\textbf{\textcolor{tagprompt}{<prompt>}}} (new request to refine or validate the answer) \texttt{\textbf{\textcolor{tagprompt}{</prompt>}}}.
Each \texttt{\textbf{\textcolor{tagprompt}{<prompt>}}} must build on the previous interaction history. Do not just repeat the same content. Let the content of \texttt{\textbf{\textcolor{tagprompt}{<prompt>}}}$\ldots$\texttt{\textbf{\textcolor{tagprompt}{</prompt>}}} evolve naturally (e.g., outline $\rightarrow$ add details $\rightarrow$ refine $\rightarrow$ check). Continue reasoning within \texttt{\textbf{\textcolor{tagthink}{<think>}}}$\ldots$\texttt{\textbf{\textcolor{tagthink}{</think>}}} and interacting within \texttt{\textbf{\textcolor{tagprompt}{<prompt>}}}$\ldots$\texttt{\textbf{\textcolor{tagprompt}{</prompt>}}} until the answer is ready.
\textbf{3.} When the answer is ready, you must output:
\texttt{\textbf{\textcolor{tagthink}{<think>}}} (final reasoning for the question) \texttt{\textbf{\textcolor{tagthink}{</think>}}}.
\texttt{\textbf{\textcolor{taganswer}{<answer>}}} (only the final answer without any explanation or analysis) \texttt{\textbf{\textcolor{taganswer}{</answer>}}}.
\noindent You must perform multi-turn interactions. Each intermediate interaction outputs:
\texttt{\textbf{\textcolor{tagthink}{<think>}}} your reasoning about a simple explanation for the question \texttt{\textbf{\textcolor{tagthink}{</think>}}}.
\texttt{\textbf{\textcolor{tagprompt}{<prompt>}}} a slightly more detailed and simple explanation of the question \texttt{\textbf{\textcolor{tagprompt}{</prompt>}}}.
\texttt{\textbf{\textcolor{tagthink}{<think>}}} your reasoning for verifying the answer \texttt{\textbf{\textcolor{tagthink}{</think>}}}.
\texttt{\textbf{\textcolor{tagprompt}{<prompt>}}} a new prompt different from the previous one, used to verify the answer \texttt{\textbf{\textcolor{tagprompt}{</prompt>}}}.
The final interaction must output:
\texttt{\textbf{\textcolor{tagthink}{<think>}}} your final reasoning for the question \texttt{\textbf{\textcolor{tagthink}{</think>}}}.
\texttt{\textbf{\textcolor{taganswer}{<answer>}}} only the final answer for the question without any explanation or analysis \texttt{\textbf{\textcolor{taganswer}{</answer>}}}.
\noindent Question: \texttt{\textbf{\textcolor{question}{question}}}.
\vspace{0.36em}
\hrule height 1pt
\end{minipage}
\end{table*}

\subsection{Markov Decision-Driven Optimization}
\label{sec:mdp}

MMDynOpt-Agent formulates the multi-turn dynamic optimization of the target MLLM by the agent as a Markov Decision Process (MDP) $\mathcal{M} = (\mathcal{S}, \mathcal{A}, \mathcal{T}, \mathcal{R}, \gamma)$, where $\mathcal{S}$ is the state space, $\mathcal{A}$ is the action space, $\mathcal{T}$ is the state transition function, $\mathcal{R}$ is the trajectory-level reward function, and $\gamma$ is the discount factor. Since the task is episodic, we set $\gamma = 1$. Each component of the MDP is as follows.

\textbf{Initialization of Environment $\mathcal{E}$ and Agent $\pi_\theta$.}
The MLLM is modeled as an environment $\mathcal{E}$, whose parameters and gradients are inaccessible to the agent $\pi_\theta$. The environment $\mathcal{E}$ internally maintains a dynamic optimization interaction history $h_t^{\mathcal{E}}$:
\begin{equation}
    h_0^{\mathcal{E}} = \emptyset, \quad h_{t+1}^{\mathcal{E}} = h_t^{\mathcal{E}} \cup \{(\textit{dop}_t,\, o_t)\}
\end{equation}
where $\textit{dop}_t$ is the dynamic optimization prompt generated 
by the agent $\pi_\theta$ at dynamic optimization round $t$, 
and $o_t$ is the reasoning response from the environment $\mathcal{E}$. 
At dynamic optimization round $t$, the environment $\mathcal{E}$ generates 
an observation based on its internal dynamic optimization history 
$h_t^{\mathcal{E}}$ and the current input:
\begin{equation}
    o_t = \mathcal{E}\!\left(\textit{dop}_t,\; q_{\text{img}} \cdot \mathbbm{1}[t{=}0],\; h_t^{\mathcal{E}}\right)
\end{equation}
where $\mathbbm{1}[\cdot]$ is the indicator function. The agent $\pi_\theta$ dynamically optimizes the reasoning of the MLLM through $\textit{dop}_t$. The history $h_t^{\mathcal{E}}$ is maintained internally by $\mathcal{E}$ and is invisible to the agent. The agent $\pi_\theta$ can only observe $o_t$ and incorporate it into its own state.

\textbf{State Space $\mathcal{S}$.}
The state is defined as the complete token sequence on the agent side. The initial state is constructed from $q$:
\begin{equation}
    s_0 = \textsc{Encode}\!\left([\text{sys};\; q_{\text{text}};\; q_{\text{img}}]\right)
\end{equation}
where $\text{sys}$ is the system instruction (see Table \ref{tab:sys-prompt-tem}), $[\,\cdot\,;\,\cdot\,]$ denotes sequence concatenation, and $\textsc{Encode}(\cdot)$ is the tokenization function. Let $|s_0| = L_0$ denote the token length of the initial state. After the $t$-th optimization round, the state is updated as 
$s_{t+1} = [s_t;\; a_t;\; o_t]$.
The state length grows monotonically with the number of rounds:
\begin{equation}
    |s_{t+1}| = |s_t| + |a_t| + |o_t| = L_0 + \sum_{k=0}^{t}\!\left(|a_k| + |o_k|\right)
\end{equation}
where $|a_t|$ and $|o_t|$ are the token lengths of the action and observation of the agent $\pi_\theta$ at the optimization round $t$, respectively.

\textbf{Markov Property.}
The state $s_t = [s_0;\; a_0;\; o_0;\; \ldots;\; a_{t-1};\; o_{t-1}]$. 
That is, $s_t$ encodes complete history from $s_0$. Therefore, for any $t$:
\begin{equation}
    P(s_{t+1} \mid s_t, a_t) = P([s_t;\, a_t;\, o_t] \mid s_t, a_t) = P(o_t \mid s_t, a_t)
\end{equation}
Since $o_t = \mathcal{E}(\textit{dop}_t, q_{\text{img}} \cdot \mathbbm{1}[t{=}0], h_t^{\mathcal{E}})$ and $h_t^{\mathcal{E}}$ can be fully reconstructed from the historical $\textit{dop}$ and $o$ sequences within $s_t$, the distribution of $o_t$ depends only on $s_t$ and $a_t$, requiring no additional history. This guarantees the Markov property of the MDP:
\begin{equation}
    P(s_{t+1} \mid s_0, a_0, o_0, \ldots, s_t, a_t) = P(s_{t+1} \mid s_t, a_t)
\end{equation}

\textbf{Action Space $\mathcal{A}$.}
The agent $\pi_\theta$ autoregressively samples actions given state $s_t$. Let $a_t = (w_1, w_2, \ldots, w_{|a_t|})$ be the token sequence of the action; the generation process is token-by-token sampling:
\begin{equation}
    \pi_\theta(a_t \mid s_t) = \prod_{j=1}^{|a_t|} \pi_\theta(w_j \mid s_t, w_{1:j-1})
\end{equation}
where $w_{1:j-1} = (w_1, \ldots, w_{j-1})$ is the prefix tokens already generated.
The agent chooses between two semantic actions each round:
\begin{equation}
    a_t = \begin{cases}
        (\text{think}_t,\; \textit{dop}_t) & \text{steering action} \\
        (\text{think}_t,\; \hat{y}) & \text{answering action}
    \end{cases}
\end{equation}
where $\text{think}_t$ is the internal reasoning chain of the agent $\pi_\theta$, $\textit{dop}_t$ is the dynamic optimization prompt which is sent to the environment $\mathcal{E}$, and $\hat{y}$ is the final predicted answer output by the agent $\pi_\theta$. 

\textbf{State Transition Function $\mathcal{T}$.}
When the agent $\pi_\theta$ executes a steering action, the environment $\mathcal{E}$ returns observation $o_t$, and the new state $s_{t+1}$ is constructed via deterministic concatenation:
\begin{equation}
    \mathcal{T}(s_{t+1} \mid s_t, a_t) = \delta\!\left(s_{t+1} = [s_t;\; a_t;\; o_t]\right)
\end{equation}
where $\delta(\cdot)$ is Dirac delta function, indicating the next state is deterministically obtained from current state and action. The stochasticity of $\mathcal{E}$ is unobservable to $\pi_\theta$ and subsumed in $o_t$. An answering action terminates dynamic optimization with no state transition.

\textbf{Termination Condition.}
The dynamic optimization between $\pi_\theta$ and $\mathcal{E}$ terminates when either of the conditions is met:
\begin{equation}
    \textit{done}(s_t, a_t) = \mathbbm{1}[a_t \text{ is answering}] \lor \mathbbm{1}[t = T]
\end{equation}
where $T$ is the preset maximum number of dynamic optimization rounds of $\pi_\theta$. If $T$ rounds are reached without generating an answering action, the dynamic optimization is forcibly terminated.

\textbf{Trajectory Probability.}
Under the above Markov Decision Process, the joint probability of a complete dynamic optimization trajectory $\tau = (s_0, a_0, o_0, \ldots, s_{T_\tau}, a_{T_\tau})$ is factorized as follows:
\begin{equation}
    P(\tau \mid \pi_\theta, \mathcal{E}) = \prod_{t=0}^{T_\tau} \pi_\theta(a_t \mid s_t) \cdot \prod_{t=0}^{T_\tau - 1} P(o_t \mid s_t, a_t)
\end{equation}
where $T_\tau \leq T$ is the actual number of dynamic optimization rounds in trajectory $\tau$. The term $\prod_{t=0}^{T_\tau} \pi_\theta(a_t \mid s_t)$ represents the cumulative decision probability of the agent policy over the entire trajectory, while $\prod_{t=0}^{T_\tau - 1} P(o_t \mid s_t, a_t)$ captures the cumulative transition probability of $\mathcal{E}$ across rounds. This factorization decouples the learnable policy $\pi_\theta$ from the environment $\mathcal{E}$ at the probabilistic level, enabling policy optimization to compute gradients solely with respect to $\pi_\theta$ without requiring an explicit model of the environment $\mathcal{E}$.

\par\noindent\textbf{Proposition 1.} \textit{Adaptive multi-turn dynamic optimization prompts can better enhance the reasoning performance of MLLMs.}\vspace{-3mm}
\begin{proof}
We provide experimental results in Section~\ref{sec:exp_main} and theoretical proofs in Appendix~\ref{app:proof1}.
\end{proof}

\subsection{Prism Reward-Guided Policy Optimization}
\label{sec:prism}

MMDynOpt-Agent introduces a trajectory-level Prism Reward $R(\tau)$ with end-to-end reinforcement learning for the policy optimization.

\textbf{Prism Reward $R(\tau)$.}
Prism Reward decomposes trajectory evaluation into three dimensions: format compliance, answer correctness, and budget efficiency, combined via hierarchical gating.

\textit{(i) Format Reward $r_f$.}
Let the agent produce $T_\tau + 1$ action blocks in trajectory $\tau$. For each intermediate action block $a_t$ ($t < T_\tau$), a reward $\alpha_{\text{mid}}$ is granted if it strictly matches the \texttt{<think>...</think>\allowbreak<prompt>...</prompt>} format; for the terminal action block $a_{T_\tau}$, rewards $\alpha_{\text{tag}}$, $\alpha_{\text{ne}}$, and $\alpha_{\text{full}}$ are assigned based on the presence of the \texttt{<answer>} tag, non-emptiness of the answer content, and complete format conformity, respectively. The format reward is defined as:
\begin{equation}
\begin{split}
    r_f(\tau) = \sum_{t=0}^{T_\tau - 1} \alpha_{\text{mid}} \cdot \mathbbm{1}[\textit{fmt}(a_t)] + \alpha_{\text{tag}} \cdot \mathbbm{1}[\textit{tag}(a_{T_\tau})] \\
    + \alpha_{\text{ne}} \cdot \mathbbm{1}[\textit{ne}(a_{T_\tau})] + \alpha_{\text{full}} \cdot \mathbbm{1}[\textit{full}(a_{T_\tau})]
\end{split}
\end{equation}
where $\textit{fmt}(\cdot)$, $\textit{tag}(\cdot)$, $\textit{ne}(\cdot)$, and $\textit{full}(\cdot)$ are predicates for intermediate-round format integrity, answer tag presence, answer content non-emptiness, and terminal-round format completeness, respectively.

\textit{(ii) Answer Reward $r_a$.}
The answer reward measures the match between the predicted answer $\hat{y}$ and the ground-truth answer set $\mathcal{G} = \{g_1, \ldots, g_{|\mathcal{G}|}\}$ via token-level F1 score. Let $\mathcal{W}_{\hat{y}}$ and $\mathcal{W}_g$ represent the normalized token sets of $\hat{y}$ and $g$, respectively:
\begin{equation}
    r_a(\tau) = \max_{g \in \mathcal{G}}\, \frac{2\,|\mathcal{W}_{\hat{y}} \cap \mathcal{W}_g|}{|\mathcal{W}_{\hat{y}}| + |\mathcal{W}_g|}
\end{equation}

\textit{(iii) Budget Efficiency Reward $r_b$.}
The reward $r_b$ encourages $\pi_\theta$ to minimize cost with $\mathcal{E}$ while maintaining answer quality:
\begin{equation}
    c_{\text{turn}} = \min\!\left(\frac{n_{\text{call}}}{N_{\text{max}}},\,1\right), c_{\text{token}} = \frac{1}{2}\left[\min\!\left(\frac{l_{\text{in}}}{L_{\text{in}}^{\max}},\,1\right) + \min\!\left(\frac{l_{\text{out}}}{L_{\text{out}}^{\max}},\,1\right)\right]
\end{equation}
where $n_{\text{call}}$ is the number of calls to $\mathcal{E}$, $N_{\text{max}}$ is the maximum turn budget, $l_{\text{in}}$ and $l_{\text{out}}$ are cumulative tokens sent to and received from $\mathcal{E}$, and $L_{\text{in}}^{\max}$ and $L_{\text{out}}^{\max}$ are the corresponding normalization bounds:
\begin{equation}
    r_b(\tau) = 1 - \tfrac{1}{2}\,c_{\text{turn}} - \tfrac{1}{2}\,c_{\text{token}}
\end{equation}

\textit{(iv) Hierarchical Gating.}
The three dimensions are composed into the trajectory-level reward via a hierarchical gating mechanism:
\begin{equation}
    R(\tau) = \begin{cases}
        -1 + r_f & \text{if } r_f < 1 \\
        -1 + r_f + r_a + \lambda_b \cdot r_a \cdot r_b \cdot \mathbbm{1}[r_a > \eta] & \text{if } r_f = 1
    \end{cases}
\end{equation}
where $\lambda_b$ is the budget efficiency weight and $\eta$ is the answer quality threshold. Answer and budget signals are masked when $r_f < 1$; budget efficiency reward is activated only when $r_f = 1$ and $r_a > \eta$.

\textbf{Dynamic Optimization Policy Learning.}
Based on the Prism Reward $R(\tau)$, the agent $\pi_\theta$ learns the optimization policy via GRPO.

\textit{(i) Group-Relative Advantage Estimation.}
For each input $q$, $\pi_\theta$ performs $G$ independent multi-turn dynamic optimizations with $\mathcal{E}$, yielding trajectories $\{\tau_1, \ldots, \tau_G\}$, each corresponding to a distinct optimization path. The group-relative advantage is defined as:
\begin{equation}
    \hat{A}_i = \frac{R(\tau_i) - \mu_G}{\sigma_G + \epsilon}, \mu_G = \frac{1}{G}\sum_{j=1}^{G} R(\tau_j), \sigma_G = \sqrt{\frac{1}{G}\sum_{j=1}^{G}(R(\tau_j) - \mu_G)^2}
\end{equation}
where $\epsilon$ is a numerical stability constant. $\hat{A}_i > 0$ indicates the dynamic optimization of trajectory $\tau_i$ outperforms the group average.

\textit{(ii) Learning Objective.}
The agent-side token sequence in multi-turn dynamic optimization contains tokens from both $\pi_\theta$ and $\mathcal{E}$. A token attribution mask $m_j^{(i)} \in \{0, 1\}$ is defined where $m_j^{(i)} = 1$ iff token $j$ in $\tau_i$ is from $\pi_\theta$. The per-token clipped surrogate loss is:
\begin{equation}
    L_j^{(i)} = \min\!\left(\rho_j^{(i)}\, \hat{A}_i,\;\; \text{clip}\!\left(\rho_j^{(i)},\, 1{-}\epsilon_l,\, 1{+}\epsilon_h\right) \hat{A}_i\right)
\end{equation}
The dynamic optimization policy learning objective of the agent is:
\begin{equation}
    \mathcal{J}(\theta) = -\mathbb{E}_{q \sim \mathcal{D}}\, \mathbb{E}_{\tau_i \sim \pi_{\theta_{\text{old}}} \otimes \mathcal{E}} \left[\frac{\sum_{j}\, m_j^{(i)} \cdot L_j^{(i)}}{\sum_{j}\, m_j^{(i)}}\right]
\end{equation}
where $\rho_j^{(i)} = \pi_\theta(w_j^{(i)} \mid s^{(i)}, w_{1:j-1}^{(i)}) \,/\, \pi_{\theta_{\text{old}}}(w_j^{(i)} \mid s^{(i)}, w_{1:j-1}^{(i)})$ is the importance sampling ratio, $\epsilon_l$ and $\epsilon_h$ are clipping parameters, and $\pi_{\theta_{\text{old}}} \otimes \mathcal{E}$ denotes trajectories generated via alternating optimization between $\pi_{\theta_{\text{old}}}$ and $\mathcal{E}$. The mask $m_j^{(i)}$ restricts gradients to $\textit{dop}_t$, $\text{think}_t$, and $\hat{y}$ tokens from $\pi_\theta$, excluding observations $o_t$ from $\mathcal{E}$.

\par\noindent\textbf{Proposition 2.} \textit{Reinforcement learning with Prism Reward enables the agent to learn a dynamic optimization policy for enhancing the reasoning of MLLMs while maintaining low reasoning overhead.}\vspace{-3mm}
\begin{proof}
We provide experimental results in Section~\ref{sec:exp_ablation} and theoretical proofs in Appendix~\ref{app:proof2}.
\end{proof}

\subsection{Transferable Optimization Framework}
\label{sec:transfer}

MMDynOpt-Agent learns a dynamic optimization policy with cross-model transferability, enabling it to adapt to a variety of MLLMs.

\textbf{Environment Agnosticism.}
The agent $\pi_\theta$ interacts with the target MLLM $\mathcal{E}$ only through a standardized interface, by sending $\textit{dop}_t$ and receiving $o_t$, without relying on the parameters or architecture of $\mathcal{E}$. Define the compatible environment set as:
\begin{equation}
    \mathbb{E} = \left\{\mathcal{E}' \mid \mathcal{E}': \mathcal{X}_{\text{text}} \times \mathcal{X}_{\text{img}} \rightarrow \mathcal{Y}_{\text{text}}\right\}
\end{equation}
where $\mathcal{X}_{\text{text}}$, $\mathcal{X}_{\text{img}}$, and $\mathcal{Y}_{\text{text}}$ are the text input, image input, and text output spaces, respectively. Any MLLM $\mathcal{E}'$ satisfying this standardized interface can serve as a drop-in replacement for $\mathcal{E}$ without modifying the agent $\pi_\theta$.
\textbf{Transfer Condition.}
Let $\pi_\theta^*$ be the policy optimized on the training environment $\mathcal{E}_{\text{train}}$. When transferring to $\mathcal{E}' \in \mathbb{E}$, the MDP changes only in the state transition function:
\begin{equation}
    o_t' = \mathcal{E}'\!\left(\textit{dop}_t,\, q_{\text{img}} \cdot \mathbbm{1}[t{=}0],\, h_t^{\mathcal{E}'}\right)
\end{equation}
\begin{equation}
    \mathcal{T}_{\mathcal{E}'}(s_{t+1} \mid s_t, a_t) = \delta\!\left(s_{t+1} = [s_t;\; a_t;\; o_t']\right)
\end{equation}
That is, $\mathcal{S}$, $\mathcal{A}$, and $\pi_\theta^*$ remain unchanged; only observations come from $\mathcal{E}'$, equivalent to replacing the MDP transition kernel.

\textbf{Dynamic Optimization Performance.}
On an MLLM $\mathcal{E}'$, the dynamic optimization effect of $\pi_\theta^*$ on its reasoning is defined as:
\begin{equation}
    \mathcal{P}_{\mathrm{dyn}}(\pi_\theta^*, \mathcal{E}') = \mathbb{E}_{q \sim \mathcal{D}_{\text{test}}} \left[R(\tau' \mid \pi_\theta^*, \mathcal{E}', q)\right]
\end{equation}
where $\mathcal{D}_{\text{test}}$ is the test data distribution and $\tau'$ is the trajectory produced by the interaction between the agent $\pi_\theta^*$ and $\mathcal{E}'$.

\textbf{Transfer Efficiency.}
Different capabilities of $\mathcal{E}'$ may lead to different dynamic optimization efficiencies. The efficiency ratio is:
\begin{equation}
    \eta_{\text{eff}}(\mathcal{E}', \mathcal{E}_{\text{train}}) =
    \frac{\mathbb{E}_{q}\!\left[T_\tau(\pi_\theta^*, \mathcal{E}_{\text{train}})\right]}
         {\mathbb{E}_{q}\!\left[T_\tau(\pi_\theta^*, \mathcal{E}')\right]}
\end{equation}
where $T_\tau(\pi_\theta^*, \mathcal{E})$ denotes the average number of dynamic optimization rounds. $\eta_{\text{eff}} > 1$ indicates that fewer rounds are required after transferring to a stronger $\mathcal{E}'$, while $\eta_{\text{eff}} < 1$ indicates the opposite.

\textbf{Reasoning Procedure.}
During inference, $\pi_\theta^*$ is frozen and performs dynamic optimization on any $\mathcal{E}' \in \mathbb{E}$ following the MDP. At each round, $\pi_\theta^*$ executes a steering action to further optimize the reasoning of $\mathcal{E}'$ through $\textit{dop}_t$, or an answering action to conclude and output $\hat{y}$. No fine-tuning of $\pi_\theta^*$ is required; only the interface of $\mathcal{E}'$ needs to be specified. The number of dynamic optimization rounds is adaptively determined based on question difficulty.

\par\noindent\textbf{Proposition 3.} \textit{The dynamic optimization policy can be effectively transferred to different MLLMs and enhance reasoning performance.}\vspace{-3mm}
\begin{proof}
We provide experimental results in Section~\ref{sec:exp_transfer} and theoretical proofs in Appendix~\ref{app:proof3}.
\end{proof}

\definecolor{grpA}{rgb}{0.9412,0.9804,0.9490}
\definecolor{grpB}{rgb}{0.9333,0.9608,0.9922}
\definecolor{grpC}{rgb}{0.9686,0.9529,0.9922}

\definecolor{grpA}{rgb}{0.9647,0.9882,0.9765}
\definecolor{grpB}{rgb}{0.9490,0.9647,0.9980}
\definecolor{grpC}{rgb}{0.9216,0.9020,0.9804}
\definecolor{grpA}{rgb}{0.97, 0.995, 0.98}

\definecolor{grpB}{rgb}{0.97, 0.99, 0.995}

\definecolor{grpC}{rgb}{0.985, 0.97, 0.995}

\begin{table*}[t]
\centering
{\fontsize{8}{8}\selectfont}
\caption{Main results of MMDynOpt-Agent (ours) and baselines on nine datasets of medicine (MedXpertQA: MedX, OmniMed: OMed, PMC-VQA: PMC), finance (FinChart-Bench: FinC, FinMME: FinM, Sujet-Finance-QA-Vision-100k: FinQA), and general science (GeoQA, SciQA, Multi). Self-Consistency is S-CON. Self-Correct is S-COR. Best values are in \textbf{bold}. All values are in \%.}
\label{tab:main-results}
\vspace{-0.6em}
\begin{tblr}{colspec={Q[l,1cm]Q[l,0.8cm]|Q[c,1.03cm]Q[c,1.03cm]Q[c,1.03cm]|Q[c,1.03cm]Q[c,1.03cm]Q[c,1.03cm]Q[c,1.03cm]Q[c,1.03cm]Q[c,1.03cm]Q[c,1.03cm]Q[c,1.03cm]Q[c,1.1cm]|Q[c,1.2cm]},
  colsep=2pt,rowsep=1pt,stretch=1.0,
  row{1}={bg=white, font=\bfseries\fontsize{8}{8}\selectfont},
  cell{1}{3}={c=3}{halign=c},
  cell{1}{6}={c=9}{halign=c},
  row{2}={bg=white, font=\fontsize{8}{8}\selectfont},
  row{3-22}={font=\fontsize{8}{8}\selectfont},
  cell{3}{1}={r=2}{},
  cell{5}{1}={r=2}{},
  cell{7}{1}={r=2}{},
  cell{9}{1}={r=2}{},
  cell{11}{1}={r=2}{},
  cell{13}{1}={r=2}{},
  cell{15}{1}={r=2}{},
  cell{17}{1}={r=2}{},
  cell{19}{1}={r=2}{},
  cell{21}{1}={r=2}{},
  column{3-5}={bg=grpA},column{6-14}={bg=grpB},column{15}={bg=grpC},
  row{1-2}={bg=white},hline{1,Z}={1.0pt},hline{3}={0.6pt}}
\textbf{Dataset} & \textbf{Metric} & \textit{\textbf{Qwen2.5-VL-3B}} & & & \textit{\textbf{Qwen3-VL-32B}} & & & & & & & & & \\
 &  & Baseline & SFT & GRPO & Baseline & CoT & OPRO & GEPA & BON & S-CON & S-COR & ToT & MARS & \textbf{Ours} \\
\textbf{MedX} & \textbf{EM} & 0.52{\scriptsize$\pm$0.45} & 25.26{\scriptsize$\pm$2.51} & 5.21{\scriptsize$\pm$2.39} & 38.02{\scriptsize$\pm$2.51} & 39.32{\scriptsize$\pm$2.51} & 36.98{\scriptsize$\pm$3.93} & 29.17{\scriptsize$\pm$1.97} & 37.76{\scriptsize$\pm$1.19} & 40.89{\scriptsize$\pm$0.91} & 39.58{\scriptsize$\pm$2.74} & 33.86{\scriptsize$\pm$3.16} & 29.95{\scriptsize$\pm$22.04} & \textbf{43.75{\scriptsize$\pm$13.35}} \\
 & \textbf{F1} & 3.83{\scriptsize$\pm$1.59} & 34.59{\scriptsize$\pm$3.58} & 9.13{\scriptsize$\pm$1.96} & 49.09{\scriptsize$\pm$1.75} & 49.03{\scriptsize$\pm$3.08} & 46.80{\scriptsize$\pm$2.50} & 41.92{\scriptsize$\pm$1.75} & 48.51{\scriptsize$\pm$0.43} & 50.34{\scriptsize$\pm$0.68} & 49.23{\scriptsize$\pm$1.60} & 44.89{\scriptsize$\pm$1.56} & 42.15{\scriptsize$\pm$15.96} & \textbf{52.73{\scriptsize$\pm$10.71}} \\
\textbf{OMed} & \textbf{EM} & 2.08{\scriptsize$\pm$0.45} & 39.84{\scriptsize$\pm$2.82} & 17.71{\scriptsize$\pm$4.31} & 67.97{\scriptsize$\pm$2.82} & 67.71{\scriptsize$\pm$1.97} & 77.08{\scriptsize$\pm$0.90} & 65.37{\scriptsize$\pm$3.85} & 73.44{\scriptsize$\pm$2.06} & 77.60{\scriptsize$\pm$1.81} & 67.19{\scriptsize$\pm$0.78} & 80.47{\scriptsize$\pm$0.78} & 33.08{\scriptsize$\pm$2.96} & \textbf{84.64{\scriptsize$\pm$5.97}} \\
 & \textbf{F1} & 10.88{\scriptsize$\pm$3.65} & 42.63{\scriptsize$\pm$3.26} & 18.98{\scriptsize$\pm$4.23} & 77.25{\scriptsize$\pm$1.82} & 79.03{\scriptsize$\pm$0.48} & 80.51{\scriptsize$\pm$0.29} & 76.83{\scriptsize$\pm$1.60} & 81.43{\scriptsize$\pm$0.89} & 83.82{\scriptsize$\pm$1.50} & 76.74{\scriptsize$\pm$1.06} & 83.24{\scriptsize$\pm$0.31} & 58.83{\scriptsize$\pm$1.44} & \textbf{88.77{\scriptsize$\pm$4.63}} \\
\textbf{PMC} & \textbf{EM} & 4.43{\scriptsize$\pm$0.45} & 39.58{\scriptsize$\pm$1.97} & 19.79{\scriptsize$\pm$3.52} & 69.79{\scriptsize$\pm$3.16} & 64.84{\scriptsize$\pm$0.78} & 67.45{\scriptsize$\pm$2.51} & 59.63{\scriptsize$\pm$2.51} & 67.71{\scriptsize$\pm$2.25} & 71.62{\scriptsize$\pm$1.97} & 61.46{\scriptsize$\pm$3.69} & 71.61{\scriptsize$\pm$3.16} & 26.56{\scriptsize$\pm$6.67} & \textbf{78.12{\scriptsize$\pm$5.90}} \\
 & \textbf{F1} & 12.35{\scriptsize$\pm$1.53} & 50.32{\scriptsize$\pm$0.40} & 26.26{\scriptsize$\pm$4.20} & 76.91{\scriptsize$\pm$2.35} & 74.19{\scriptsize$\pm$0.89} & 73.41{\scriptsize$\pm$2.94} & 70.00{\scriptsize$\pm$1.82} & 74.76{\scriptsize$\pm$1.03} & 77.48{\scriptsize$\pm$1.06} & 70.64{\scriptsize$\pm$2.99} & 76.24{\scriptsize$\pm$2.67} & 41.44{\scriptsize$\pm$4.21} & \textbf{82.93{\scriptsize$\pm$4.06}} \\
\textbf{FinC} & \textbf{EM} & 13.02{\scriptsize$\pm$1.80} & 17.45{\scriptsize$\pm$3.16} & 17.97{\scriptsize$\pm$3.41} & 57.29{\scriptsize$\pm$1.19} & 56.77{\scriptsize$\pm$1.96} & 57.29{\scriptsize$\pm$0.90} & 53.65{\scriptsize$\pm$0.46} & 57.29{\scriptsize$\pm$0.45} & 57.03{\scriptsize$\pm$0.78} & 56.25{\scriptsize$\pm$1.56} & 55.73{\scriptsize$\pm$0.90} & 57.29{\scriptsize$\pm$0.45} & \textbf{57.81{\scriptsize$\pm$0.78}} \\
 & \textbf{F1} & 21.51{\scriptsize$\pm$3.90} & 26.28{\scriptsize$\pm$2.41} & 29.00{\scriptsize$\pm$2.41} & 71.11{\scriptsize$\pm$0.79} & 70.67{\scriptsize$\pm$1.36} & 71.55{\scriptsize$\pm$0.57} & 69.02{\scriptsize$\pm$0.24} & 70.61{\scriptsize$\pm$0.31} & 71.90{\scriptsize$\pm$0.44} & 69.54{\scriptsize$\pm$1.12} & 68.82{\scriptsize$\pm$0.70} & 68.42{\scriptsize$\pm$2.91} & \textbf{72.50{\scriptsize$\pm$1.40}} \\
\textbf{FinM} & \textbf{EM} & 14.32{\scriptsize$\pm$1.97} & 14.06{\scriptsize$\pm$0.78} & 15.88{\scriptsize$\pm$1.20} & 47.14{\scriptsize$\pm$0.45} & 47.14{\scriptsize$\pm$1.20} & 46.62{\scriptsize$\pm$1.20} & 40.89{\scriptsize$\pm$0.91} & 46.35{\scriptsize$\pm$1.97} & 48.96{\scriptsize$\pm$0.45} & 45.83{\scriptsize$\pm$1.97} & 47.66{\scriptsize$\pm$0.78} & 42.97{\scriptsize$\pm$3.13} & \textbf{51.56{\scriptsize$\pm$2.82}} \\
 & \textbf{F1} & 20.64{\scriptsize$\pm$1.73} & 20.87{\scriptsize$\pm$1.75} & 23.26{\scriptsize$\pm$1.80} & 57.83{\scriptsize$\pm$0.81} & 57.77{\scriptsize$\pm$0.87} & 57.12{\scriptsize$\pm$0.54} & 52.59{\scriptsize$\pm$0.57} & 57.96{\scriptsize$\pm$2.05} & 59.74{\scriptsize$\pm$0.14} & 56.75{\scriptsize$\pm$2.61} & 59.11{\scriptsize$\pm$1.90} & 54.46{\scriptsize$\pm$3.58} & \textbf{65.61{\scriptsize$\pm$1.95}} \\
\textbf{FinQA} & \textbf{EM} & 0.78{\scriptsize$\pm$0.78} & 2.34{\scriptsize$\pm$1.56} & 2.34{\scriptsize$\pm$0.78} & 3.12{\scriptsize$\pm$0.00} & 3.12{\scriptsize$\pm$0.00} & 3.65{\scriptsize$\pm$0.46} & 3.12{\scriptsize$\pm$0.00} & 3.65{\scriptsize$\pm$0.46} & 3.12{\scriptsize$\pm$0.00} & 3.12{\scriptsize$\pm$0.00} & 3.38{\scriptsize$\pm$0.46} & 3.91{\scriptsize$\pm$0.00} & \textbf{4.69{\scriptsize$\pm$0.00}} \\
 & \textbf{F1} & 7.13{\scriptsize$\pm$2.06} & 19.80{\scriptsize$\pm$2.14} & 12.91{\scriptsize$\pm$2.76} & 34.25{\scriptsize$\pm$0.17} & 34.02{\scriptsize$\pm$0.36} & 33.90{\scriptsize$\pm$0.83} & 32.47{\scriptsize$\pm$1.06} & 34.13{\scriptsize$\pm$0.40} & 34.36{\scriptsize$\pm$0.36} & 19.78{\scriptsize$\pm$0.33} & 32.69{\scriptsize$\pm$0.74} & 34.86{\scriptsize$\pm$0.16} & \textbf{39.83{\scriptsize$\pm$2.80}} \\
\textbf{GeoQA} & \textbf{EM} & 3.91{\scriptsize$\pm$0.79} & 9.38{\scriptsize$\pm$2.35} & 4.43{\scriptsize$\pm$0.90} & 35.42{\scriptsize$\pm$2.96} & 36.72{\scriptsize$\pm$3.12} & 35.68{\scriptsize$\pm$2.39} & 38.28{\scriptsize$\pm$0.78} & 37.76{\scriptsize$\pm$2.51} & 38.02{\scriptsize$\pm$1.80} & 37.50{\scriptsize$\pm$1.56} & 36.72{\scriptsize$\pm$1.35} & 29.69{\scriptsize$\pm$4.14} & \textbf{39.58{\scriptsize$\pm$4.44}} \\
 & \textbf{F1} & 9.28{\scriptsize$\pm$0.49} & 13.41{\scriptsize$\pm$2.34} & 9.71{\scriptsize$\pm$2.74} & 59.44{\scriptsize$\pm$2.65} & 60.44{\scriptsize$\pm$1.20} & 59.52{\scriptsize$\pm$0.99} & 61.21{\scriptsize$\pm$1.12} & 57.98{\scriptsize$\pm$5.64} & 62.61{\scriptsize$\pm$0.82} & 59.33{\scriptsize$\pm$2.67} & 58.61{\scriptsize$\pm$0.51} & 59.33{\scriptsize$\pm$3.62} & \textbf{68.52{\scriptsize$\pm$3.37}} \\
\textbf{SciQA} & \textbf{EM} & 1.82{\scriptsize$\pm$0.45} & 52.86{\scriptsize$\pm$3.16} & 23.18{\scriptsize$\pm$4.77} & 77.34{\scriptsize$\pm$1.56} & 78.39{\scriptsize$\pm$1.63} & 81.25{\scriptsize$\pm$3.41} & 73.18{\scriptsize$\pm$0.90} & 77.34{\scriptsize$\pm$1.56} & 78.39{\scriptsize$\pm$3.93} & 78.90{\scriptsize$\pm$1.36} & 89.58{\scriptsize$\pm$1.19} & 73.96{\scriptsize$\pm$8.64} & \textbf{96.61{\scriptsize$\pm$1.80}} \\
 & \textbf{F1} & 13.61{\scriptsize$\pm$2.94} & 64.69{\scriptsize$\pm$1.21} & 28.01{\scriptsize$\pm$3.01} & 87.59{\scriptsize$\pm$0.56} & 87.99{\scriptsize$\pm$0.88} & 87.09{\scriptsize$\pm$2.44} & 84.81{\scriptsize$\pm$0.48} & 86.82{\scriptsize$\pm$0.21} & 87.85{\scriptsize$\pm$2.95} & 87.78{\scriptsize$\pm$1.25} & 95.38{\scriptsize$\pm$0.40} & 85.69{\scriptsize$\pm$3.48} & \textbf{97.77{\scriptsize$\pm$0.81}} \\
\textbf{Multi} & \textbf{EM} & 10.68{\scriptsize$\pm$4.30} & 30.99{\scriptsize$\pm$7.26} & 24.48{\scriptsize$\pm$3.85} & 42.71{\scriptsize$\pm$2.51} & 48.18{\scriptsize$\pm$2.25} & 45.84{\scriptsize$\pm$1.81} & 36.98{\scriptsize$\pm$1.62} & 44.27{\scriptsize$\pm$2.96} & 46.09{\scriptsize$\pm$0.79} & 43.75{\scriptsize$\pm$2.71} & 56.51{\scriptsize$\pm$1.19} & 34.38{\scriptsize$\pm$1.57} & \textbf{66.15{\scriptsize$\pm$4.71}} \\
 & \textbf{F1} & 17.71{\scriptsize$\pm$2.92} & 35.28{\scriptsize$\pm$6.81} & 28.55{\scriptsize$\pm$2.06} & 61.38{\scriptsize$\pm$1.62} & 64.51{\scriptsize$\pm$1.11} & 62.79{\scriptsize$\pm$2.04} & 58.54{\scriptsize$\pm$1.44} & 62.65{\scriptsize$\pm$2.07} & 63.24{\scriptsize$\pm$0.70} & 61.09{\scriptsize$\pm$2.07} & 67.66{\scriptsize$\pm$0.81} & 56.20{\scriptsize$\pm$0.40} & \textbf{72.24{\scriptsize$\pm$2.67}} \\
\textbf{Overall} & \textbf{EM} & 5.73{\scriptsize$\pm$1.27} & 25.75{\scriptsize$\pm$2.84} & 14.55{\scriptsize$\pm$2.79} & 48.76{\scriptsize$\pm$1.91} & 49.13{\scriptsize$\pm$1.71} & 50.20{\scriptsize$\pm$1.94} & 44.47{\scriptsize$\pm$1.44} & 49.51{\scriptsize$\pm$1.71} & 51.30{\scriptsize$\pm$1.38} & 48.18{\scriptsize$\pm$1.82} & 52.84{\scriptsize$\pm$1.44} & 36.86{\scriptsize$\pm$5.51} & \textbf{58.10{\scriptsize$\pm$4.42}} \\
 & \textbf{F1} & 12.99{\scriptsize$\pm$2.31} & 34.21{\scriptsize$\pm$2.65} & 20.65{\scriptsize$\pm$2.80} & 63.87{\scriptsize$\pm$1.39} & 64.18{\scriptsize$\pm$1.14} & 63.63{\scriptsize$\pm$1.46} & 60.82{\scriptsize$\pm$1.12} & 63.87{\scriptsize$\pm$1.45} & 65.71{\scriptsize$\pm$0.96} & 61.21{\scriptsize$\pm$1.74} & 65.18{\scriptsize$\pm$1.07} & 55.71{\scriptsize$\pm$3.97} & \textbf{71.21{\scriptsize$\pm$3.60}} \\
\end{tblr}
\end{table*}

\section{Experiments}
In this section, we report the experimental setup, results, and analysis. We answer the following research questions (RQ): 
\textbf{RQ1:} Can MMDynOpt-Agent enhance multimodal reasoning of MLLMs and outperform baselines? \textbf{RQ2:} Does the learned dynamic optimization policy generalize to out-of-distribution datasets? \textbf{RQ3:} Can the MMDynOpt-Agent transfer to other MLLMs across architectures and scales? \textbf{RQ4:} How effective are MMDynOpt-Agent’s components, and what does comparative analysis reveal? \textbf{RQ5:} How efficient is MMDynOpt-Agent in a performance-cost trade-off?

\subsection{Experimental Setup}
\label{sec:exp_setup}

\textbf{Datasets.}
We evaluate MMDynOpt-Agent on 15 multimodal benchmarks across \textbf{medicine}, \textbf{finance}, and \textbf{general science}, including
\textbf{PMC-VQA}~\cite{zhang2023pmc}, 
\textbf{MedXpertQA}~\cite{zuo2025medxpertqa},
\textbf{PathVQA}~\cite{he2020pathvqa}, 
\textbf{VQA-RAD}~\cite{lau2018dataset}, 
\textbf{OmniMedVQA}~\cite{hu2024omnimedvqa}, 
\textbf{FinChart-Bench}~\cite{shu2025finchart},  
\textbf{Sujet-Finance-QA-Vision-100k}~\cite{Sujet-Finance-QA-Vision-100k}, 
\textbf{FinMME}~\cite{luo2025finmme},
\textbf{VisFinEval}~\cite{liu2025visfineval}, 
\textbf{MathVista}~\cite{lu2023mathvista}, 
\textbf{Multi}~\cite{zhu2025multi}, 
\textbf{GeoQA}~\cite{chen2021geoqa}, \textbf{SeePhys}~\cite{xiang2025seephys}, 
\textbf{AI2D}~\cite{kembhavi2016diagram},
and \textbf{ScienceQA}~\cite{lu2022learn}. 
More details are illustrated in Appendix~\ref{app:datasets}.

\textbf{Evaluation Metrics.}
We evaluate MMDynOpt-Agent and the baselines on these metrics: Exact Match (\textbf{EM}), token-level F1 score (\textbf{F1}), semantic similarity (\textbf{SSim}), and Total Token per F1 (\textbf{FPT}). More details of evaluation metrics are illustrated in Appendix~\ref{app:metrics}.

\textbf{Baselines.}
We mainly compare MMDynOpt-Agent with the baseline LLMs~\cite{team2024gemma2,wang2025internvl3,deshmukh2025nvidia}, SFT, GRPO~\cite{guo2025deepseek}, CoT~\cite{mitra2024compositional}, GEPA~\cite{agrawal2025gepa}, OPRO~\cite{yang2023large}, Self-Correct~\cite{welleck2022generating}, ToT~\cite{yao2023tree}, Best-of-N (BoN)~\cite{brown2024large}, Self-Consistency~\cite{wang2022self}, and MARS~\cite{zhang2025mars}. More details are in Appendix~\ref{app:baselines}.

\textbf{Implementation Details.}
We utilize Qwen2.5-VL-3B-Instruct \cite{bai2025qwen25vltechnicalreport} as the agent, which is trained with the target MLLMs (Qwen3-VL-32B-Instruct \cite{bai2025qwen3}). All experiments are done on a server with eight NVIDIA H100 GPUs (80GB). More details are in Appendix~\ref{app:implementation}.

\definecolor{grpA}{rgb}{0.97, 0.995, 0.98}

\definecolor{grpB}{rgb}{0.97, 0.99, 0.995}

\definecolor{grpC}{rgb}{0.985, 0.97, 0.995}

\begin{table*}[t]
\centering
{\fontsize{8}{8}\selectfont}
\caption{Comparison results of the baselines and MMDynOpt-Agent (ours) on the six out-of-distribution (OOD) public multimodal datasets, which includes MathVista (Math), PathVQA (Path), SeePhys (SPhys), VisFinEval (VisF), AI2D, and VQA-RAD (RAD).}
\label{tab:ood}
\vspace{-0.6em}
\begin{tblr}{colspec={Q[l,1cm]Q[l,0.8cm]|Q[c,1.03cm]Q[c,1.03cm]Q[c,1.03cm]|Q[c,1.03cm]Q[c,1.03cm]Q[c,1.03cm]Q[c,1.03cm]Q[c,1.03cm]Q[c,1.03cm]Q[c,1.03cm]Q[c,1.03cm]Q[c,1.1cm]|Q[c,1.2cm]},
  colsep=2pt,rowsep=1pt,stretch=1.0,
  row{1}={bg=white, font=\bfseries\fontsize{8}{8}\selectfont},
  cell{1}{3}={c=3}{halign=c},
  cell{1}{6}={c=9}{halign=c},
  row{2}={bg=white, font=\fontsize{8}{8}\selectfont},
  row{3-16}={font=\fontsize{8}{8}\selectfont},
  cell{3}{1}={r=2}{},
  cell{5}{1}={r=2}{},
  cell{7}{1}={r=2}{},
  cell{9}{1}={r=2}{},
  cell{11}{1}={r=2}{},
  cell{13}{1}={r=2}{},
  cell{15}{1}={r=2}{},
  column{3-5}={bg=grpA},column{6-14}={bg=grpB},column{15}={bg=grpC},
  row{1-2}={bg=white},hline{1,Z}={1.0pt},hline{3}={0.6pt}}
\textbf{Dataset} & \textbf{Metric} & \textit{\textbf{Qwen2.5-VL-3B}} & & & \textit{\textbf{Qwen3-VL-32B}} & & & & & & & & & \\
 &  & Baseline & SFT & GRPO & Baseline & CoT & OPRO & GEPA & BON & S-CON & S-COR & ToT & MARS & \textbf{Ours} \\
\textbf{Math} & \textbf{EM} & 10.16{\scriptsize$\pm$1.35} & 13.02{\scriptsize$\pm$2.51} & 14.32{\scriptsize$\pm$1.19} & 63.02{\scriptsize$\pm$0.90} & 61.98{\scriptsize$\pm$2.51} & 64.58{\scriptsize$\pm$1.63} & 60.42{\scriptsize$\pm$1.97} & 62.76{\scriptsize$\pm$2.51} & 66.93{\scriptsize$\pm$1.20} & 61.20{\scriptsize$\pm$1.19} & 56.77{\scriptsize$\pm$2.96} & 70.05{\scriptsize$\pm$0.90} & \textbf{70.31{\scriptsize$\pm$2.34}} \\
 & \textbf{F1} & 12.85{\scriptsize$\pm$2.31} & 16.29{\scriptsize$\pm$3.75} & 16.14{\scriptsize$\pm$0.83} & 69.83{\scriptsize$\pm$0.30} & 68.59{\scriptsize$\pm$1.72} & 70.47{\scriptsize$\pm$1.46} & 67.04{\scriptsize$\pm$2.32} & 69.38{\scriptsize$\pm$2.52} & 73.24{\scriptsize$\pm$1.06} & 68.30{\scriptsize$\pm$1.19} & 65.19{\scriptsize$\pm$2.36} & 74.50{\scriptsize$\pm$1.50} & \textbf{75.39{\scriptsize$\pm$4.57}} \\
\textbf{Path} & \textbf{EM} & 5.21{\scriptsize$\pm$1.19} & 50.52{\scriptsize$\pm$2.74} & 19.27{\scriptsize$\pm$4.44} & 71.61{\scriptsize$\pm$4.30} & 67.71{\scriptsize$\pm$1.62} & 75.52{\scriptsize$\pm$1.96} & 67.71{\scriptsize$\pm$3.16} & 73.96{\scriptsize$\pm$0.45} & 76.04{\scriptsize$\pm$0.90} & 70.05{\scriptsize$\pm$0.45} & 70.83{\scriptsize$\pm$2.39} & 69.01{\scriptsize$\pm$1.96} & \textbf{81.51{\scriptsize$\pm$3.16}} \\
 & \textbf{F1} & 11.99{\scriptsize$\pm$0.98} & 50.65{\scriptsize$\pm$2.77} & 19.53{\scriptsize$\pm$4.35} & 73.96{\scriptsize$\pm$2.71} & 72.83{\scriptsize$\pm$2.09} & 77.60{\scriptsize$\pm$1.97} & 74.31{\scriptsize$\pm$1.77} & 76.39{\scriptsize$\pm$0.75} & 76.91{\scriptsize$\pm$0.60} & 73.18{\scriptsize$\pm$0.26} & 75.52{\scriptsize$\pm$1.38} & 69.18{\scriptsize$\pm$2.24} & \textbf{81.51{\scriptsize$\pm$3.16}} \\
\textbf{SPhys} & \textbf{EM} & 0.00{\scriptsize$\pm$0.00} & 0.26{\scriptsize$\pm$0.45} & 0.78{\scriptsize$\pm$0.00} & 1.30{\scriptsize$\pm$0.45} & 1.82{\scriptsize$\pm$0.90} & 1.56{\scriptsize$\pm$0.78} & 1.82{\scriptsize$\pm$0.45} & 1.04{\scriptsize$\pm$0.45} & 1.56{\scriptsize$\pm$0.00} & 0.78{\scriptsize$\pm$0.78} & 1.04{\scriptsize$\pm$0.45} & 2.34{\scriptsize$\pm$1.36} & \textbf{3.91{\scriptsize$\pm$0.78}} \\
 & \textbf{F1} & 7.32{\scriptsize$\pm$0.84} & 15.41{\scriptsize$\pm$2.01} & 15.80{\scriptsize$\pm$0.60} & 44.04{\scriptsize$\pm$0.91} & 44.52{\scriptsize$\pm$0.52} & 41.96{\scriptsize$\pm$0.64} & 41.46{\scriptsize$\pm$1.20} & 43.87{\scriptsize$\pm$1.19} & 45.64{\scriptsize$\pm$0.21} & 42.47{\scriptsize$\pm$1.87} & 41.22{\scriptsize$\pm$1.43} & 43.29{\scriptsize$\pm$0.78} & \textbf{48.87{\scriptsize$\pm$2.83}} \\
\textbf{VisF} & \textbf{EM} & 2.08{\scriptsize$\pm$0.90} & 26.04{\scriptsize$\pm$5.32} & 16.93{\scriptsize$\pm$4.30} & 61.98{\scriptsize$\pm$0.45} & 62.24{\scriptsize$\pm$3.25} & 68.75{\scriptsize$\pm$2.06} & 49.74{\scriptsize$\pm$2.51} & 63.54{\scriptsize$\pm$4.01} & 66.93{\scriptsize$\pm$0.45} & 60.94{\scriptsize$\pm$3.41} & 64.84{\scriptsize$\pm$2.07} & 51.30{\scriptsize$\pm$9.92} & \textbf{79.17{\scriptsize$\pm$3.93}} \\
 & \textbf{F1} & 11.66{\scriptsize$\pm$3.00} & 32.70{\scriptsize$\pm$3.94} & 24.20{\scriptsize$\pm$3.32} & 70.26{\scriptsize$\pm$0.63} & 71.34{\scriptsize$\pm$2.49} & 75.69{\scriptsize$\pm$1.69} & 60.60{\scriptsize$\pm$1.71} & 72.37{\scriptsize$\pm$2.91} & 74.67{\scriptsize$\pm$0.62} & 70.15{\scriptsize$\pm$2.87} & 72.68{\scriptsize$\pm$1.62} & 62.79{\scriptsize$\pm$7.24} & \textbf{81.03{\scriptsize$\pm$3.73}} \\
\textbf{AI2D} & \textbf{EM} & 11.98{\scriptsize$\pm$1.80} & 37.24{\scriptsize$\pm$1.19} & 28.39{\scriptsize$\pm$2.26} & 83.34{\scriptsize$\pm$3.16} & 82.81{\scriptsize$\pm$2.82} & 82.82{\scriptsize$\pm$2.71} & 84.90{\scriptsize$\pm$1.20} & 82.81{\scriptsize$\pm$0.78} & 83.85{\scriptsize$\pm$0.46} & 78.65{\scriptsize$\pm$0.46} & 79.17{\scriptsize$\pm$0.45} & 33.59{\scriptsize$\pm$40.02} & \textbf{92.19{\scriptsize$\pm$2.34}} \\
 & \textbf{F1} & 23.65{\scriptsize$\pm$1.79} & 42.39{\scriptsize$\pm$2.06} & 34.92{\scriptsize$\pm$1.80} & 89.70{\scriptsize$\pm$2.60} & 89.72{\scriptsize$\pm$1.58} & 89.23{\scriptsize$\pm$2.30} & 89.75{\scriptsize$\pm$0.79} & 89.67{\scriptsize$\pm$0.87} & 90.91{\scriptsize$\pm$0.13} & 87.89{\scriptsize$\pm$0.16} & 86.33{\scriptsize$\pm$0.82} & 67.06{\scriptsize$\pm$17.41} & \textbf{94.06{\scriptsize$\pm$1.43}} \\
\textbf{RAD} & \textbf{EM} & 4.95{\scriptsize$\pm$2.26} & 15.10{\scriptsize$\pm$1.20} & 10.42{\scriptsize$\pm$2.96} & 20.57{\scriptsize$\pm$0.45} & 20.57{\scriptsize$\pm$1.20} & 33.59{\scriptsize$\pm$2.07} & 33.07{\scriptsize$\pm$2.74} & 19.79{\scriptsize$\pm$2.39} & 24.22{\scriptsize$\pm$2.34} & 16.15{\scriptsize$\pm$1.97} & 11.98{\scriptsize$\pm$1.62} & 25.78{\scriptsize$\pm$7.16} & \textbf{42.19{\scriptsize$\pm$6.25}} \\
 & \textbf{F1} & 6.26{\scriptsize$\pm$2.24} & 19.54{\scriptsize$\pm$2.18} & 12.89{\scriptsize$\pm$3.19} & 35.17{\scriptsize$\pm$1.46} & 34.23{\scriptsize$\pm$0.94} & 43.70{\scriptsize$\pm$2.34} & 44.36{\scriptsize$\pm$2.74} & 35.01{\scriptsize$\pm$1.88} & 38.00{\scriptsize$\pm$1.62} & 24.78{\scriptsize$\pm$2.33} & 29.19{\scriptsize$\pm$1.12} & 36.81{\scriptsize$\pm$5.80} & \textbf{54.73{\scriptsize$\pm$6.01}} \\
\textbf{Overall} & \textbf{EM} & 5.73{\scriptsize$\pm$1.25} & 23.70{\scriptsize$\pm$2.23} & 15.02{\scriptsize$\pm$2.53} & 50.30{\scriptsize$\pm$1.62} & 49.52{\scriptsize$\pm$2.05} & 54.47{\scriptsize$\pm$1.87} & 49.61{\scriptsize$\pm$2.00} & 50.65{\scriptsize$\pm$1.76} & 53.25{\scriptsize$\pm$0.89} & 47.96{\scriptsize$\pm$1.38} & 47.44{\scriptsize$\pm$1.66} & 42.01{\scriptsize$\pm$10.22} & \textbf{61.55{\scriptsize$\pm$3.13}} \\
 & \textbf{F1} & 12.29{\scriptsize$\pm$1.86} & 29.50{\scriptsize$\pm$2.78} & 20.58{\scriptsize$\pm$2.35} & 63.83{\scriptsize$\pm$1.43} & 63.54{\scriptsize$\pm$1.56} & 66.44{\scriptsize$\pm$1.73} & 62.92{\scriptsize$\pm$1.76} & 64.45{\scriptsize$\pm$1.69} & 66.56{\scriptsize$\pm$0.71} & 61.13{\scriptsize$\pm$1.45} & 61.69{\scriptsize$\pm$1.45} & 58.94{\scriptsize$\pm$5.83} & \textbf{72.60{\scriptsize$\pm$3.62}} \\
\end{tblr}
\end{table*}

\subsection{Main Results (RQ1)}
\label{sec:exp_main}
As shown in Table~\ref{tab:main-results}, MMDynOpt-Agent achieves the best performance across nine datasets of medicine, finance, and general science. The results reveal:
\textit{(i)} Dynamic Optimization Enhances Reasoning Effectiveness.
Under the same target MLLM, MMDynOpt-Agent achieves the best overall performance without modifying the target model's parameters, demonstrating that dynamic reasoning optimization effectively enhances MLLM reasoning. Post-training methods (SFT, GRPO) on the small agent model yield improvements but remain constrained by limited capacity, further validating the agent-driven dynamic optimization paradigm.
\textit{(ii)} Adaptive Feedback Outperforms Fixed Strategies.
Structured prompting (e.g., CoT) and automatic prompt optimization (e.g., OPRO, GEPA) rely on fixed templates, while search-based test-time scaling (e.g., ToT, Self-Consistency) depends on manually designed strategies. MMDynOpt-Agent adaptively controls reasoning via reinforcement learning, leading to stronger overall performance.
\textit{(iii)} Trajectory-Level Optimization Improves Stability.
MARS exhibits high variance on multiple datasets, whereas MMDynOpt-Agent shows lower variance in most cases, suggesting that optimizing complete trajectories with global reward signals improves overall stability, though variance remains non-negligible on certain challenging datasets.

\subsection{Out-of-Distribution Generalization (RQ2)}
\label{sec:exp_ood}
As shown in Table~\ref{tab:ood}, MMDynOpt-Agent achieves the best overall performance across six out-of-distribution datasets, demonstrating remarkable generalization capability. These results provide the following insights:
\textit{(i)} The dynamic optimization policy transfers across datasets. MMDynOpt-Agent is never exposed to these datasets during training, yet still achieves the best performance, suggesting that the agent acquires transferable reasoning optimization capabilities rather than relying on dataset-specific patterns.
\textit{(ii)} Dynamic optimization yields larger gains on harder tasks. On datasets where baseline performance is relatively low, MMDynOpt-Agent delivers substantially larger improvements over the next-best method than on datasets where baselines already perform well, suggesting that multi-turn adaptive optimization becomes increasingly powerful as reasoning difficulty grows.
\textit{(iii)} Post-training methods show limited OOD generalization. SFT performs considerably worse than MMDynOpt-Agent on OOD datasets, suggesting that directly fine-tuning model parameters may induce stronger distribution dependence, whereas dynamic optimization-based reasoning enhancement exhibits stronger generalization robustness.

\begin{figure}[t]
  \centering
  \includegraphics[width=1\linewidth]{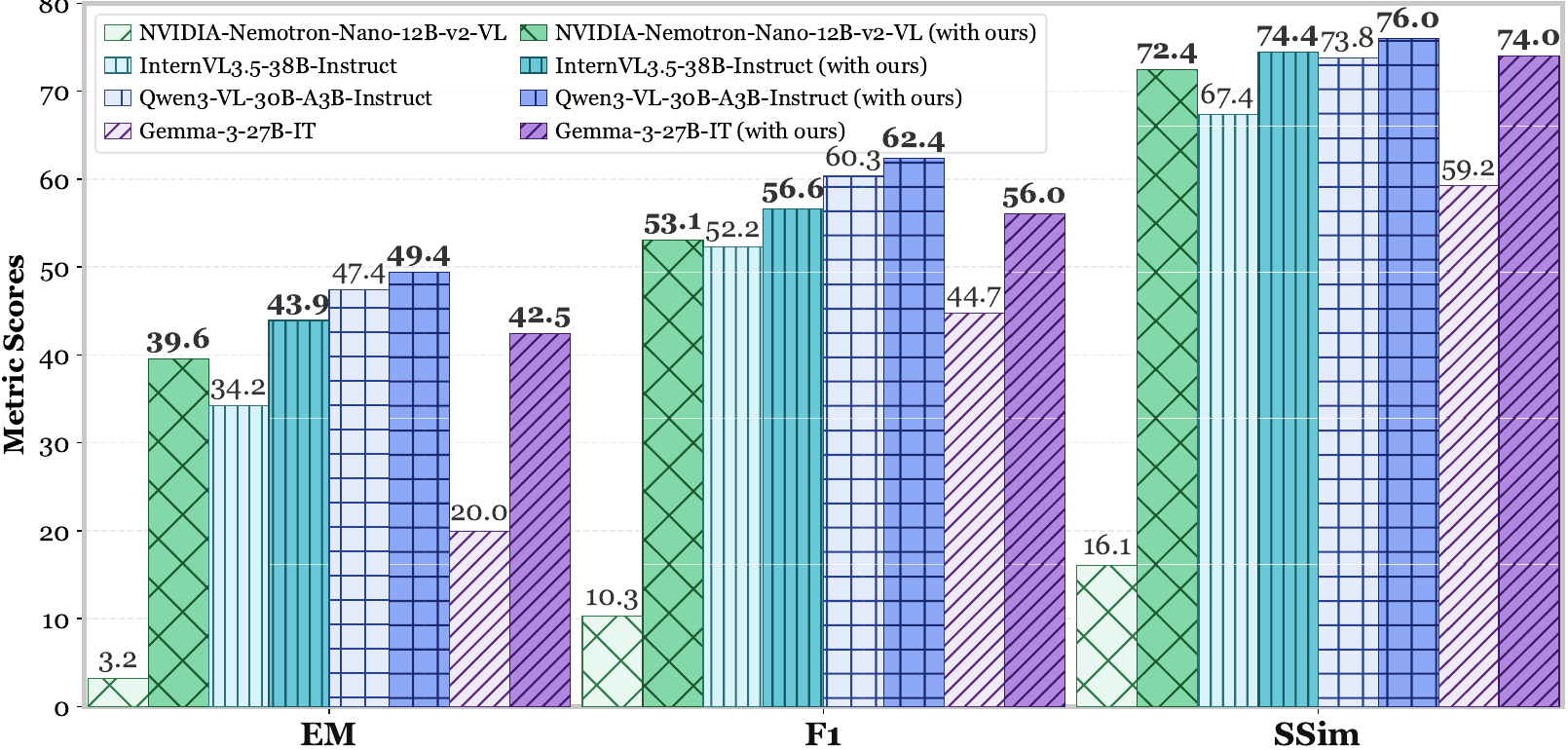}
  \vspace{-5mm}
  \caption{Comparison of average EM, F1, and SSim of 4 MLLMs with and without MMDynOpt-Agent on 9 training datasets.}
 \label{fig: four_ood_MLLM_training_dataset}
\end{figure}

\begin{figure}[t]
  \centering
  \includegraphics[width=1\linewidth]{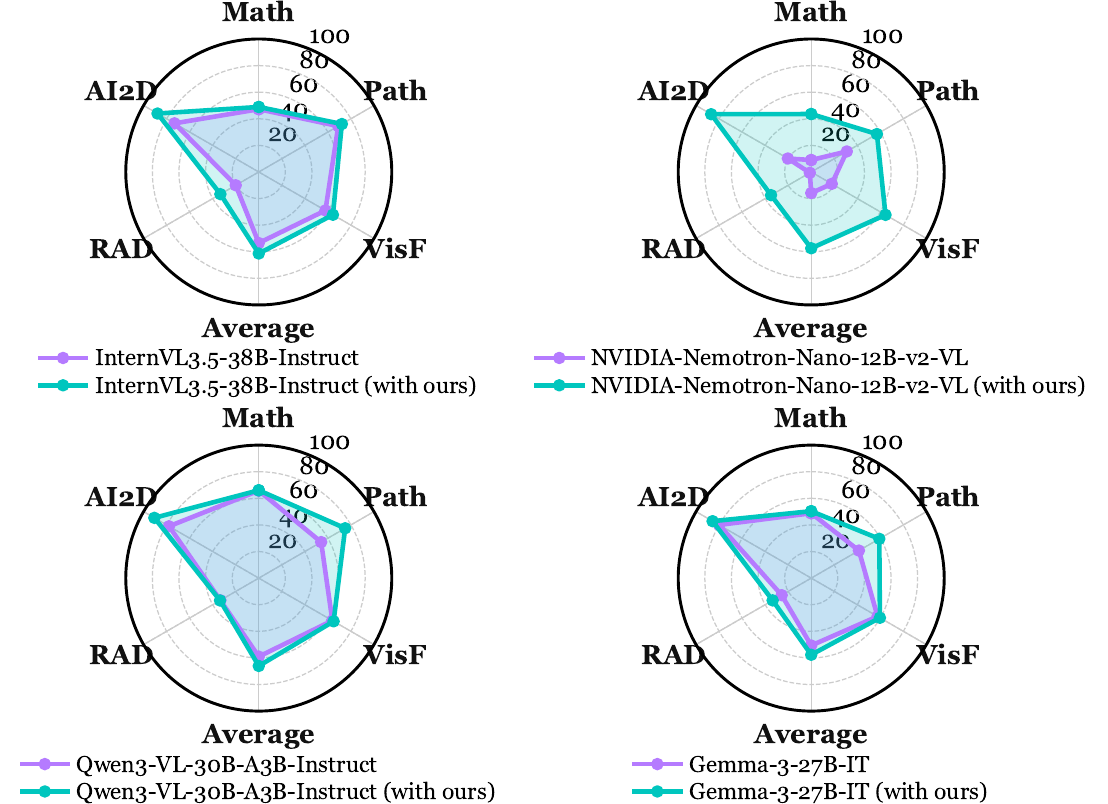}
  \vspace{-5mm}
  \caption{Comparison of F1 scores of four unseen MLLMs with and without MMDynOpt-Agent on five OOD datasets.}
 \label{fig: four_ood_MLLM_ood_dataset}
\end{figure}

\definecolor{grpA}{rgb}{0.9412,0.9804,0.9490}
\definecolor{grpB}{rgb}{0.9608,0.9765,0.9980}
\definecolor{grpC}{rgb}{0.985, 0.97, 0.995}
\begin{table*}[t]
\centering
\caption{Ablation study of MMDynOpt-Agent on six datasets across medicine, finance, and general science, including without reinforcement learning (w/o R.L.), without the trained MMDynOpt-Agent (w/o Agent), and without Prism Reward (w/o P.R.)}
\label{tab:ablation}
\vspace{-0.6em}
\begin{tblr}{colspec={Q[l,2.8cm]Q[c,1cm]Q[c,1cm]Q[c,1cm]Q[c,1cm]Q[c,1cm]Q[c,1cm]Q[c,1cm]Q[c,1cm]Q[c,1cm]Q[c,1cm]Q[c,1cm]Q[c,1cm]},
  colsep=3pt,rowsep=1pt,stretch=1.0,
  row{1}={bg=white, font=\bfseries\fontsize{8}{8}\selectfont},
  row{2}={bg=white, font=\bfseries\fontsize{8}{8}\selectfont},
  row{3-5}={font=\fontsize{8}{8}\selectfont},
  row{6}={font=\fontsize{8}{8}\selectfont, bg=grpC},
  cell{1}{2}={c=2}{halign=c},
  cell{1}{4}={c=2}{halign=c},
  cell{1}{6}={c=2}{halign=c},
  cell{1}{8}={c=2}{halign=c},
  cell{1}{10}={c=2}{halign=c},
  cell{1}{12}={c=2}{halign=c},
  hline{1,Z}={1.0pt},
  hline{2}={2}{0.4pt, leftpos=-1, rightpos=0},
  hline{2}={3}{0.4pt, leftpos=0, rightpos=-1},
  hline{2}={4}{0.4pt, leftpos=-1, rightpos=0},
  hline{2}={5}{0.4pt, leftpos=0, rightpos=-1},
  hline{2}={6}{0.4pt, leftpos=-1, rightpos=0},
  hline{2}={7}{0.4pt, leftpos=0, rightpos=-1},
  hline{2}={8}{0.4pt, leftpos=-1, rightpos=0},
  hline{2}={9}{0.4pt, leftpos=0, rightpos=-1},
  hline{2}={10}{0.4pt, leftpos=-1, rightpos=0},
  hline{2}={11}{0.4pt, leftpos=0, rightpos=-1},
  hline{2}={12}{0.4pt, leftpos=-1, rightpos=0},
  hline{2}={13}{0.4pt, leftpos=0, rightpos=-1},
  hline{3}={0.6pt}}
\textbf{Method} & \textbf{OmniMed} & & \textbf{PathVQA} & & \textbf{FinMME} & & \textbf{VisFinEval} & & \textbf{SciQA} & & \textbf{AI2D} & \\
 & \textbf{EM} & \textbf{F1} & \textbf{EM} & \textbf{F1} & \textbf{EM} & \textbf{F1} & \textbf{EM} & \textbf{F1} & \textbf{EM} & \textbf{F1} & \textbf{EM} & \textbf{F1} \\
\textit{w/o R.L.} & 8.59 & 27.23 & 2.34 & 21.88 & 21.88 & 31.19 & 4.69 & 22.30 & 2.34 & 28.39 & 10.94 & 29.04 \\
\textit{w/o Agent} & 67.97 & 77.25 & 71.61 & 73.96 & 47.14 & 57.83 & 61.98 & 70.26 & 77.34 & 87.59 & 83.34 & 89.70 \\
\textit{w/o P.R.} & 71.88 & 77.30 & 74.22 & 78.61 & 49.22 & 64.91 & 71.88 & 78.20 & 89.06 & 93.71 & 83.59 & 89.71 \\
MMDynOpt-Agent & \textbf{84.64} & \textbf{88.77} & \textbf{81.51} & \textbf{81.51} & \textbf{51.56} & \textbf{65.61} & \textbf{79.17} & \textbf{81.03} & \textbf{96.61} & \textbf{97.77} & \textbf{92.19} & \textbf{94.06} \\
\end{tblr}
\end{table*}

\begin{figure}[t]
    \centering
    \includegraphics[width=0.49\columnwidth]{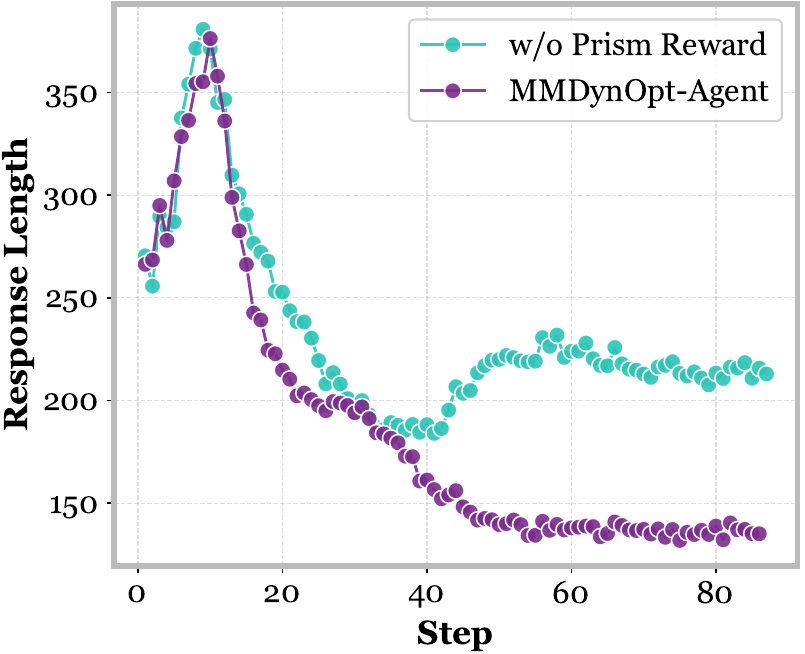}\hfill
    \includegraphics[width=0.49\columnwidth]{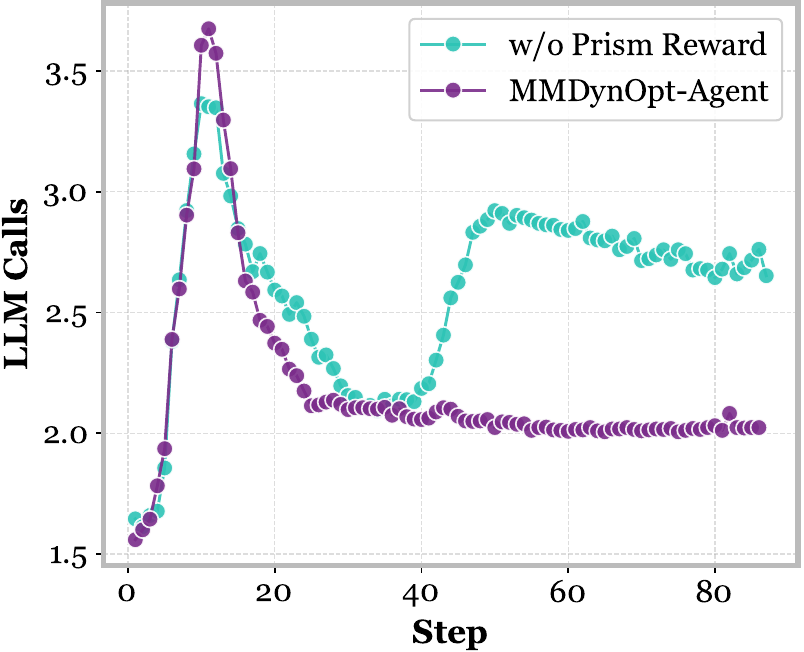}
    
    \parbox{0.49\columnwidth}{\centering\small (a) Response Length}\hfill
    \parbox{0.49\columnwidth}{\centering\small (b) LLM Calls}
    \vspace{-3mm}
    \caption{Response length and MLLM calls of the MMDynOpt-Agent with and without Prism Reward in training process.}
    \label{fig:training}
\end{figure}

\definecolor{grpA}{rgb}{0.9412,0.9804,0.9490}
\definecolor{grpB}{rgb}{0.9608,0.9765,0.9980}
\definecolor{grpC}{rgb}{0.985, 0.97, 0.995}
\begin{table}[t]
\centering
\caption{Efficiency comparison with multi-step baselines on 15 datasets, counting tokens processed by the target MLLM.}
\label{tab:efficiency}
\vspace{-0.6em}
\begin{tblr}{
  colspec={Q[l,1.56cm]Q[c,1.1cm]Q[c,1.1cm]Q[c,1.1cm]Q[c,1.16cm]Q[c,1.1cm]},
  colsep=3pt, rowsep=1pt, stretch=1.0,
  row{1}={bg=white, font=\bfseries\fontsize{8}{8}\selectfont},
  row{2}={bg=white, font=\bfseries\fontsize{8}{8}\selectfont},
  row{3-7}={font=\fontsize{8}{8}\selectfont},
  row{8}={bg=grpC, font=\fontsize{8}{8}\selectfont},
  cell{1}{2}={c=2}{halign=c},
  cell{1}{4}={c=3}{halign=c},
  hline{1,Z}={1.0pt},
  hline{2}={2}{0.4pt, leftpos=-1, rightpos=0},
  hline{2}={3}{0.4pt, leftpos=0, rightpos=-1},
  hline{2}={4}{0.4pt, leftpos=-1, rightpos=0},
  hline{2}={5}{0.4pt, leftpos=0, rightpos=0},
  hline{2}={6}{0.4pt, leftpos=0, rightpos=-1},
  hline{3}={0.6pt}
}
\textbf{Method} & \textbf{Average Metric} & & \textbf{Per-Sample Token Cost} & & \\
 & \textbf{EM} & \textbf{F1} & \textbf{Input} & \textbf{Output} & \textbf{Total} \\
BON            & 49.97 & 64.10 & 10260.30  & 6756.83  & 17017.13  \\
S-CON          & 52.08 & 66.05 & 4479.45   & 6644.52  & 11123.98  \\
S-COR          & 48.09 & 61.18 & 19691.35  & 3015.19  & 22706.54  \\
ToT            & 50.68 & 63.78 & 378531.20 & 34710.74 & 413241.93 \\
MARS           & 38.92 & 57.00 & 12684.30  & 3662.56  & 16346.86  \\
Ours & \textbf{59.48} & \textbf{71.77} & \textbf{1911.61} & \textbf{861.48} & \textbf{2773.09} \\
\end{tblr}
\end{table}

\subsection{Transferability Results and Analysis (RQ3)}
\label{sec:exp_transfer}
As shown in Figure~\ref{fig: four_ood_MLLM_training_dataset} and Figure~\ref{fig: four_ood_MLLM_ood_dataset}, we directly transfer the trained MMDynOpt-Agent to four unseen MLLMs without any fine-tuning. The results show:
\textit{(i)} The dynamic optimization policy exhibits strong cross-model generalization. The agent consistently improves performance on four unseen MLLMs, indicating that the learned optimization strategy is not tightly coupled with any specific model and generalizes well across different architectures.
\textit{(ii)} Weaker models benefit more. For models with lower baseline performance (e.g., Nemotron and Gemma), the performance gains after transfer are substantially larger than those of stronger models, suggesting that dynamic optimization provides greater improvement potential for weaker reasoning models.
\textit{(iii)} The transfer gains also hold on OOD datasets. All four models achieve consistent improvements on out-of-distribution datasets, suggesting that the agent captures more generalizable reasoning optimization capabilities rather than relying on patterns specific to individual models or datasets.

\subsection{Ablation Study (RQ4)}
\label{sec:exp_ablation}
As shown in Table~\ref{tab:ablation}, removing any component degrades MMDynOpt-Agent's performance, confirming each module's importance. Specifically:
\textit{(i)} Reinforcement learning is a foundation for dynamic optimization. Removing RL causes a substantial performance drop, indicating that an untrained agent struggles to generate effective dynamic optimization prompts. This highlights the role of end-to-end policy learning in dynamic reasoning optimization.
\textit{(ii)} The trained agent significantly enhances MLLM reasoning. Without the agent, the system reduces to direct inference with the underlying MLLM, leading to lower performance than the full system. This demonstrates that the agent is the core driver of dynamic optimization by guiding the reasoning process.
\textit{(iii)} Prism Reward improves both optimization quality and efficiency. Using Prism Reward provides more effective training signals for learning stronger optimization policies, resulting in better performance. It also encourages shorter responses and fewer MLLM calls during training, suggesting improved optimization efficiency and stability.

\begin{figure}[t]
  \centering
  \includegraphics[width=1\linewidth]{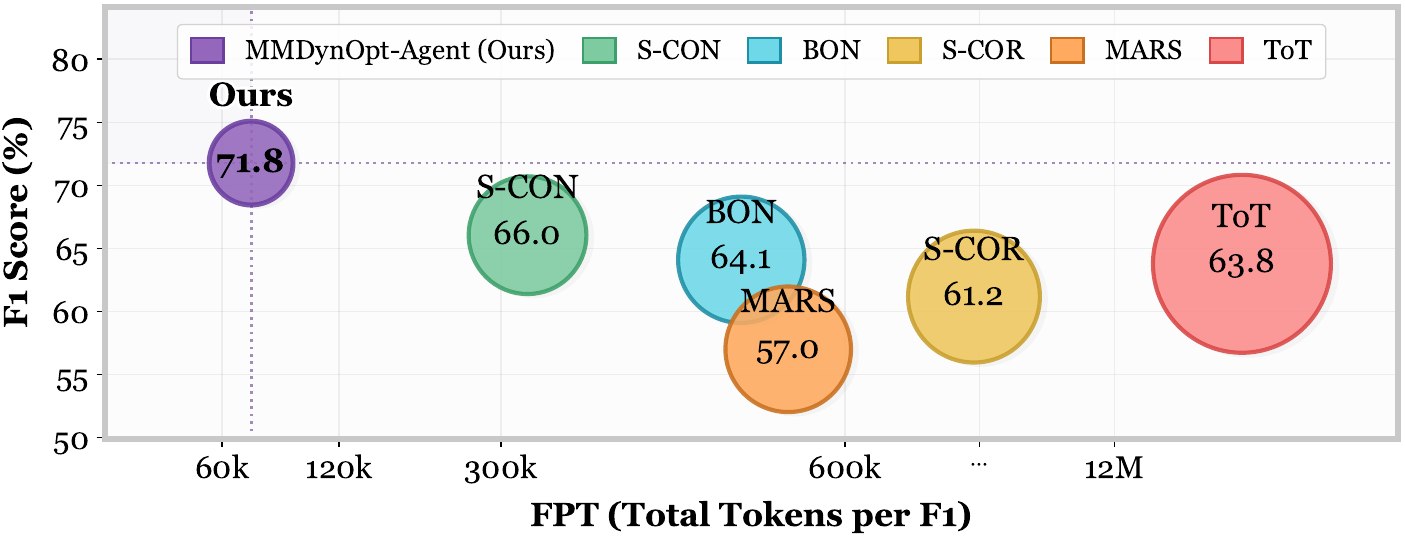}
  \vspace{-5mm}
  \caption{Comparison of performance-cost trade-off between MMDynOpt-Agent and multi-step reasoning baselines.}
 \label{fig: FPT}
\end{figure}

\subsection{Performance-Cost Efficiency (RQ5)}
\label{sec:exp_efficiency}
As shown in Table~\ref{tab:efficiency} and Figure~\ref{fig: FPT}, we compare MMDynOpt-Agent with multi-step reasoning baselines (BON, Self-Consistency, Self-Correct, ToT, MARS) in terms of performance and token cost. The results illustrate that:
\textit{(i)} MMDynOpt-Agent achieves the best results with the lowest inference cost. It attains the highest EM and F1 while using the fewest average tokens per sample, over two orders of magnitude fewer than ToT. This advantage may arise because other methods rely on fixed sampling or search depths, whereas MMDynOpt-Agent learns an adaptive turn control policy through reinforcement learning to allocate inference cost by problem difficulty and avoid unnecessary token consumption.
\textit{(ii)} The FPT metric further validates the efficiency advantage. MMDynOpt-Agent achieves a much lower FPT than all baselines, indicating that it consumes significantly fewer tokens per unit of F1 
gained. This confirms that the efficiency advantage stems from genuinely improved reasoning rather than merely shortening output length.

\section{Conclusion}
\label{conclusion}
In this work, we propose MMDynOpt-Agent, an end-to-end reinforcement learning framework for multimodal reasoning optimization. It formulates multi-turn reasoning as a Markov decision process, where a lightweight agent adaptively generates prompts to guide MLLM reasoning. Prism Reward enforces format, correctness, and budget awareness via hierarchical gating, improving the reasoning quality-cost trade-off. Policy–environment decoupling enables cross-model transfer without fine-tuning. Experiments on 15 datasets across medicine, finance, and general science show that MMDynOpt-Agent outperforms baselines in performance, generalization, transferability, and efficiency. We hope this inspires autonomous reasoning optimization in general multimodal agents.

\bibliographystyle{ACM-Reference-Format}
\bibliography{sample-base}

\newpage
\appendix

\section*{Appendix}

\section{Theoretical Proof}

\subsection{Proof of Proposition 1}
\label{app:proof1}

\par\noindent\textbf{Proposition 1.} \textit{Adaptive multi-turn dynamic optimization prompts can better enhance the reasoning performance of MLLMs.}
\vspace{-3mm}
\begin{proof}
Let $\mathcal{Y}$ be the answer semantic space equipped with metric $d: \mathcal{Y} \times \mathcal{Y} \to \mathbb{R}_{\geq 0}$ such that $(\mathcal{Y}, d)$ is a complete metric space. For multimodal input $q = (q_{\text{text}}, q_{\text{img}})$, let $y^*$ be the ground-truth answer. Without any reasoning optimization, the MLLM $\mathcal{E}$ produces an initial estimate $\hat{y}_0$, where $\hat{y}_0 \neq y^*$. At round $t$, the agent $\pi_\theta$ generates steering action $a_t = (\text{think}_t, \textit{dop}_t)$ from state $s_t$; $\mathcal{E}$ returns observation $o_t$ based on $q_{\text{img}}$ and history $h_t^{\mathcal{E}}$; the estimate updates from $\hat{y}_t$ to $\hat{y}_{t+1}$.

Let $\mathcal{B} \subseteq \mathcal{Y}$ be a closed subset containing $y^*$. Define the dynamic optimization operator $\Phi: \mathcal{B} \to \mathcal{B}$, $\Phi(\hat{y}_t) = \hat{y}_{t+1}$, representing one complete cycle of analyzing current reasoning, generating $\textit{dop}_t$, receiving $o_t$, and updating the estimate. The agent $\pi_\theta$ analyzes visual regions and semantic cues of $(q_{\text{text}}, q_{\text{img}})$ each round and guides $\mathcal{E}$ to correct prior deviations via $\textit{dop}_t$. Based on this design mechanism, it is assumed that $\Phi$ satisfies the Lipschitz contraction condition: $\exists\, 0 < \kappa < 1$,
\begin{equation}
    d(\Phi(\hat{y}),\, \Phi(\hat{y}')) \leq \kappa \cdot d(\hat{y},\, \hat{y}'), \quad \forall\, \hat{y}, \hat{y}' \in \mathcal{B},
\end{equation}
where $\kappa$ quantifies the per-round compression efficiency on reasoning deviation. It is further assumed that $\Phi$ satisfies the correctness preservation condition:
$\Phi(y^*) = y^*$,
i.e., the agent introduces no deviation when the answer is correct, so $y^*$ is a fixed point of $\Phi$. As $\mathcal{B}$ is closed in complete $(\mathcal{Y}, d)$, $(\mathcal{B}, d)$ is complete. By the Banach contraction mapping theorem, $\Phi$ has a unique fixed point on $\mathcal{B}$; since $y^*$ is one, uniqueness yields that this fixed point is $y^*$. Recursive application of the contraction gives:
\begin{equation}
    d(\hat{y}_1, y^*) = d(\Phi(\hat{y}_0), \Phi(y^*)) \leq \kappa \cdot d(\hat{y}_0, y^*),
\end{equation}
\begin{equation}
    d(\hat{y}_2, y^*) = d(\Phi(\hat{y}_1), \Phi(y^*)) \leq \kappa^2 \cdot d(\hat{y}_0, y^*).
\end{equation}
By induction, for any $t \in \{0, 1, \ldots, T\}$, the a priori error bound is:
\begin{equation}
    d(\hat{y}_t, y^*) \leq \kappa^t \cdot d(\hat{y}_0, y^*).
\end{equation}
Since $0<\kappa<1$ and $d(\hat{y}_0, y^*) > 0$, the upper bound sequence $\kappa^t \cdot d(\hat{y}_0, y^*)$ is strictly monotonically decreasing in $t$ and converges to zero, i.e., the reasoning error is controlled by a geometrically convergent strictly decreasing sequence:
\begin{equation}
    d(\hat{y}_0, y^*) > \kappa \cdot d(\hat{y}_0, y^*) > \kappa^2 \cdot d(\hat{y}_0, y^*) > \cdots > \kappa^T \cdot d(\hat{y}_0, y^*) > 0.
\end{equation}
The a posteriori error bound of Picard iteration gives:
\begin{equation}
    d(\hat{y}_{t+1}, y^*) \leq \frac{\kappa}{1-\kappa} \cdot d(\hat{y}_{t+1}, \hat{y}_t),
\end{equation}
and the inter-round change satisfies:
\begin{equation}
    d(\hat{y}_{t+1}, \hat{y}_t) = d(\Phi(\hat{y}_t), \Phi(\hat{y}_{t-1})) \leq \kappa \cdot d(\hat{y}_t, \hat{y}_{t-1}) \leq \kappa^t \cdot d(\hat{y}_1, \hat{y}_0).
\end{equation}

Consider one-shot reasoning optimization methods (e.g., structured prompting, automatic prompt optimization, test-time scaling, multi-agent systems) that apply a single fixed transformation to $\mathcal{E}$ without iterative refinement. Let $\hat{y}_{\text{static}}$ be the resulting answer. For problem instances on which one-shot methods leave non-zero residual error, let $\xi \in (0, 1)$ denote the corresponding error retention factor:
\begin{equation}
    d(\hat{y}_{\text{static}}, y^*) = \xi \cdot d(\hat{y}_0, y^*),
\end{equation}
where $\xi > 0$ reflects the non-trivial case in which a one-shot method does not directly reach the ground-truth answer. For such instances, after $T_\tau$ adaptive rounds, $d(\hat{y}_{T_\tau}, y^*) \leq \kappa^{T_\tau} d(\hat{y}_0, y^*)$. When $T_\tau > \log \xi / \log \kappa$, we have $\kappa^{T_\tau} < \xi$. Since $\log \xi < 0$ and $\log \kappa < 0$, this is satisfied in finitely many rounds, giving:
\begin{equation}
    d(\hat{y}_{T_\tau}, y^*) \leq \kappa^{T_\tau} \cdot d(\hat{y}_0, y^*) < \xi \cdot d(\hat{y}_0, y^*) = d(\hat{y}_{\text{static}}, y^*).
\end{equation}
Thus, for any instance on which one-shot optimization leaves non-zero residual error, sufficiently many adaptive rounds yield strictly smaller reasoning error.

To extend the above deterministic analysis to the probabilistic setting, consider the randomness of problem instances $q$ sampled from data distribution $\mathcal{D}$. For $q$ drawn from $\mathcal{D}$, the initial error $d(\hat{y}_0, y^*)$ and the one-shot error retention factor $\xi$ are random variables. Let $\delta > 0$ be a decision threshold, $P_{\text{acc}}^{(\text{static})} \triangleq P(d(\hat{y}_{\text{static}}, y^*) < \delta)$, and $P_{\text{acc}}^{(T)} \triangleq P(d(\hat{y}_{T_\tau}, y^*) < \delta)$. Define the event
\begin{equation}
A = \{\xi > 0,\; T_\tau > \log \xi / \log \kappa\}.
\end{equation}
For all instances in $A$, the deterministic analysis gives $d(\hat{y}_{T_\tau}, y^*) < d(\hat{y}_{\text{static}}, y^*)$, which implies
\begin{equation}
    \{d(\hat{y}_{\text{static}}, y^*) < \delta\} \cap A \subseteq \{d(\hat{y}_{T_\tau}, y^*) < \delta\} \cap A.
\end{equation}
Hence,
\begin{equation}
P(d(\hat{y}_{T_\tau}, y^*) < \delta \mid A) \geq P(d(\hat{y}_{\text{static}}, y^*) < \delta \mid A).
\end{equation}
Moreover, when
\begin{equation}
d(\hat{y}_{\text{static}}, y^*) \in [\delta,\, \delta\xi/\kappa^{T_\tau}),
\end{equation}
we have
\begin{equation}
    d(\hat{y}_{T_\tau}, y^*) \leq (\kappa^{T_\tau}/\xi) \cdot d(\hat{y}_{\text{static}}, y^*) < \delta,
\end{equation}
that is, one-shot optimization fails but multi-turn optimization succeeds. If the data distribution $\mathcal{D}$ assigns positive probability to such instances, then the multi-turn method achieves a strict accuracy gain on a non-zero-measure subset of the data, implying a strict overall improvement under this distributional condition:
\begin{equation}
    P_{\text{acc}}^{(T)} > P_{\text{acc}}^{(\text{static})}.
\end{equation}

The agent $\pi_\theta$ autonomously chooses between steering and answering at each round; the interaction count $T_\tau \leq T$ adapts to problem difficulty, terminating when inter-round change drops below $\varepsilon > 0$ or round $T$ is reached:
\begin{equation}
    T_\tau = \min\!\left(\min\{t \geq 1 : d(\hat{y}_t, \hat{y}_{t-1}) < \varepsilon\},\; T\right).
\end{equation}
Since $d(\hat{y}_t, \hat{y}_{t-1}) \leq \kappa^{t-1} d(\hat{y}_1, \hat{y}_0)$, the stopping-round upper bound is:
\begin{equation}
    T_\tau \leq \min\!\left(1 + \left\lceil \frac{\log(d(\hat{y}_1, \hat{y}_0)/\varepsilon)}{\log(1/\kappa)} \right\rceil,\; T\right).
\end{equation}
Difficult problems yield large $d(\hat{y}_1, \hat{y}_0)$ and thus larger $T_\tau$; easy problems yield small $d(\hat{y}_1, \hat{y}_0)$ and earlier termination. At stopping, the a posteriori error bound gives $d(\hat{y}_{T_\tau}, y^*) \leq \frac{\kappa}{1-\kappa} \varepsilon$. Setting $\varepsilon = \delta(1-\kappa)/\kappa$ for any target precision $\delta > 0$ guarantees $d(\hat{y}_{T_\tau}, y^*) < \delta$, with required rounds:
\begin{equation}
    T_\tau \leq \min\!\left(1 + \left\lceil \frac{\log\!\left(\kappa \cdot d(\hat{y}_1, \hat{y}_0)\, /\, (\delta(1-\kappa))\right)}{\log(1/\kappa)} \right\rceil,\; T\right).
\end{equation}

In summary, adaptive multi-turn dynamic optimization contracts reasoning error by $\kappa < 1$ per round to $\kappa^{T_\tau} d(\hat{y}_0, y^*)$ after $T_\tau$ rounds; for instances on which one-shot methods leave non-zero residual error, multi-turn optimization attains strictly smaller error after finitely many rounds; under the above distributional condition, this also yields strictly higher accuracy; the stopping rounds scale logarithmically with problem difficulty.
\end{proof}

\subsection{Proof of Proposition 2}
\label{app:proof2}

\par\noindent\textbf{Proposition 2.} \textit{Reinforcement learning with Prism Reward enables the agent to learn a dynamic optimization policy for enhancing the reasoning of MLLMs while maintaining low reasoning overhead.}
\vspace{-3mm}
\begin{proof}
Let the Markov decision process be
\[
\mathcal{M}=(\mathcal{S},\mathcal{A},\mathcal{T},\mathcal{R},\gamma),
\]
where the agent $\pi_\theta$ interacts with the environment $\mathcal{E}$ to generate trajectory $\tau$, and $\gamma=1$. Prism Reward combines the format reward $r_f$, the answer reward $r_a\in[0,1]$, and the budget efficiency reward $r_b\in[0,1]$ through hierarchical gating, where $\lambda_b>0$ is the efficiency weight and $\eta\in(0,1)$ is the quality threshold:
\[
R(\tau)=
\begin{cases}
-1+r_f, & r_f<1,\\[1mm]
-1+r_f+r_a+\lambda_b r_a r_b\,\mathbbm{1}[r_a>\eta], & r_f=1.
\end{cases}
\]
Since $T_\tau\le T$ and all reward components are bounded, there exist finite constants $R_{\min},R_{\max}$ such that
\[
R(\tau)\in [R_{\min},R_{\max}].
\]

Define the reasoning quality objective as
\[
J_q(\theta)=\mathbb{E}_{\tau\sim\pi_\theta}[r_a(\tau)],
\]
and define the reasoning overhead as
\[
\Omega(\theta)=\mathbb{E}_{\tau\sim\pi_\theta}[1-r_b(\tau)].
\]
Then
\[
1-\Omega(\theta)=\mathbb{E}_{\tau\sim\pi_\theta}[r_b(\tau)]
\]
represents the average efficiency level. When $r_f=1$ and $r_a>\eta$, Prism Reward reduces to
\[
R(\tau)=r_a(1+\lambda_b r_b).
\]
In this activated-gating regime, define the sample-level scalarization function
\[
u(r_a,r_b)=r_a(1+\lambda_b r_b).
\]
Its partial derivatives are
\[
\frac{\partial u}{\partial r_a}=1+\lambda_b r_b>0,\qquad
\frac{\partial u}{\partial r_b}=\lambda_b r_a>0.
\]
Hence, $u$ is strictly increasing in both $r_a$ and $r_b$. Therefore, in the regime where $r_f=1$ and $r_a>\eta$, if a local policy change satisfies
\[
\Delta r_a\ge 0,\qquad \Delta r_b\ge 0,
\]
with at least one strict inequality, then
\[
\Delta u>0.
\]
This shows that Prism Reward provides a consistent joint optimization direction for reasoning quality and reasoning efficiency in the high-quality regime, while $\lambda_b$ controls their trade-off: larger $\lambda_b$ biases optimization more toward efficiency, whereas smaller $\lambda_b$ biases optimization more toward quality.

Furthermore, the hierarchical gating mechanism of Prism Reward decomposes training into progressive subtasks. When $r_f<1$,
\[
R(\tau)=-1+r_f,
\]
and policy updates mainly learn the correct output format. When $r_f=1$ but $r_a\le \eta$,
\[
R(\tau)=r_a,
\]
and policy updates mainly improve answer correctness. When $r_f=1$ and $r_a>\eta$,
\[
R(\tau)=r_a(1+\lambda_b r_b),
\]
the policy starts to jointly optimize quality and efficiency. Therefore, the training process follows a progressive scheme from format compliance to answer correctness and finally to a balance between quality and efficiency. Since only part of the reward components is activated in each stage, the effective reward range in each stage is typically smaller than the global reward range, which helps reduce gradient-estimation variance and improve training stability.

Next, we analyze the policy learning process. Let the parameter at iteration $k$ be $\theta_k$, let the policy-gradient estimator be $\hat g(\theta_k)$, and let the true gradient be
\[
g(\theta_k)=\nabla_\theta J(\theta_k).
\]
Then the parameter update is written as
\[
\theta_{k+1}
=
\theta_k+\alpha_k\bigl[g(\theta_k)+\xi_k\bigr],
\]
where $\alpha_k$ is the learning rate and $\xi_k=\hat g(\theta_k)-g(\theta_k)$ is the stochastic gradient noise. Assume
\[
\sum_{k=0}^{\infty}\alpha_k=\infty,\qquad
\sum_{k=0}^{\infty}\alpha_k^2<\infty,
\]
and
\[
\mathbb{E}[\xi_k\mid \theta_k]=0,\qquad
\mathbb{E}[\|\xi_k\|^2]\le \sigma^2<\infty.
\]
Also assume that the objective function $J(\theta)$ satisfies standard smoothness conditions over the reachable parameter region, and that finite step sizes together with the clipped objective jointly ensure stable updates.

Since $R(\tau)$ is bounded, if
\[
\mathbb{E}\!\left[\|\nabla_\theta\log\pi_\theta(\tau)\|^2\right]<\infty,
\]
then under REINFORCE or policy-gradient estimation, the gradient noise satisfies
\[
\mathbb{E}[\|\xi_k\|^2]
\;\lesssim\;
\frac{R_{\mathrm{range}}^2\,
\mathbb{E}\!\left[\|\nabla_\theta\log\pi_\theta(\tau)\|^2\right]}{G}
<\infty,
\]
where $G$ is the number of sampled trajectories in each group. Moreover, group-relative advantage normalization and hierarchical gating constrain the fluctuation of the advantage terms and the effective reward range, respectively, and thus jointly help satisfy the bounded-variance condition.

Under the above standard conditions, the stochastic gradient iteration satisfies the first-order stationarity convergence result:
\[
\min_{0\le k\le K}\mathbb{E}\!\left[\|\nabla_\theta J(\theta_k)\|^2\right]\to 0,
\qquad K\to\infty.
\]
Therefore, there exists a parameter subsequence $\{\theta_{k_j}\}$ such that
\[
\mathbb{E}\!\left[\|\nabla_\theta J(\theta_{k_j})\|^2\right]\to 0.
\]
Let $\theta^\star$ denote its limit point. Then $\theta^\star$ is a first-order stationary point of the objective function $J$.

Moreover, since in the high-quality gating regime,
\[
\frac{\partial u}{\partial r_a}>0,\qquad
\frac{\partial u}{\partial r_b}>0,
\]
there does not exist, in a neighborhood of the stationary point, any local parameter perturbation satisfying
\[
\Delta r_a\ge 0,\qquad \Delta \Omega\le 0,
\]
with at least one strict inequality, that can still continue to strictly improve the objective value. Equivalently, the learned policy reaches the following local balance in that neighborhood: answer quality can no longer be significantly improved without increasing reasoning overhead, and reasoning overhead can no longer be significantly reduced without degrading answer quality. Hence, the resulting policy does not merely pursue longer reasoning traces or more interaction rounds, but instead forms an adaptive dynamic optimization behavior between quality and overhead.

In summary, Prism Reward forms a strictly monotone joint objective over reasoning quality and reasoning efficiency in the high-quality gating regime; its hierarchical gating mechanism decomposes training into progressive subtasks and helps reduce the variance of gradient estimation; under standard smoothness, unbiasedness, bounded-variance, and stable-update conditions, policy-gradient optimization converges to a first-order stationary point of the objective, thereby learning a dynamic optimization policy that balances reasoning quality and reasoning overhead.
\end{proof}

\subsection{Proof of Proposition 3}
\label{app:proof3}

\par\noindent\textbf{Proposition 3.} \textit{The dynamic optimization policy can be effectively transferred to different MLLMs and enhance reasoning performance.}
\vspace{-3mm}
\begin{proof}
Let $\pi_\theta^*$ be the dynamic optimization policy trained on environment $\mathcal{E}_{\mathrm{train}}$. The interaction between the agent $\pi_\theta^*$ and an MLLM is conducted through a unified interface: sending dynamic optimization prompts $\textit{dop}_t$ and receiving observations $o_t$, without depending on the internal parameters or specific architecture of environment $\mathcal{E}$. Define the class of compatible environments as
\[
\mathbb{E}
=
\left\{
\mathcal{E}'
\mid
\mathcal{E}':
\mathcal{X}_{\text{text}}\times \mathcal{X}_{\text{img}}
\to
\mathcal{Y}_{\text{text}}
\right\}.
\]
For any $\mathcal{E}'\in\mathbb{E}$, the input-output interface between the agent and the environment remains unchanged. Therefore, the policy $\pi_\theta^*$ can be directly applied to this environment without modifying its parameters. After transfer, the state space $\mathcal{S}$, action space $\mathcal{A}$, and reward function $\mathcal{R}$ remain unchanged; only the state transition kernel changes:
\[
\begin{aligned}
\mathcal{T}_{\mathcal{E}'}(s_{t+1}\mid s_t,a_t)
&=
\delta\!\left(s_{t+1}=[s_t;\,a_t;\,o_t']\right),\\
o_t'
&=
\mathcal{E}'\!\left(
\textit{dop}_t,\;
q_{\text{img}}\cdot \mathbbm{1}[t=0],\;
h_t^{\mathcal{E}'}
\right).
\end{aligned}
\]

Let $(\mathcal{Y},d)$ be a complete metric space over the answer space, and let $\mathcal{B}\subseteq\mathcal{Y}$ be a closed subset containing the ground-truth answer $y^*$. For any environment $\mathcal{E}$, define the dynamic optimization operator
\[
\Phi_{\mathcal{E}}:\mathcal{B}\to\mathcal{B},
\qquad
\Phi_{\mathcal{E}}(\hat y_t)=\hat y_{t+1}.
\]
As in Proposition~1, assume that the operator $\Phi_{\mathcal{E}_{\mathrm{train}}}$ on the training environment satisfies the contraction condition, i.e., there exists
\[
\kappa_{\mathcal{E}_{\mathrm{train}}}\in(0,1)
\]
such that
\[
d\!\left(
\Phi_{\mathcal{E}_{\mathrm{train}}}(\hat y),
\Phi_{\mathcal{E}_{\mathrm{train}}}(\hat y')
\right)
\le
\kappa_{\mathcal{E}_{\mathrm{train}}}\,
d(\hat y,\hat y'),
\qquad
\forall \hat y,\hat y'\in\mathcal{B},
\]
and assume that it satisfies the correctness-preservation condition
\[
\Phi_{\mathcal{E}_{\mathrm{train}}}(y^*)=y^*.
\]

Model the environment transfer $\mathcal{E}_{\mathrm{train}}\to\mathcal{E}'$ as a perturbation of the dynamic optimization operator
\[
\Phi_{\mathcal{E}_{\mathrm{train}}}
\to
\Phi_{\mathcal{E}'}.
\]
Further assume that $\Phi_{\mathcal{E}'}$ also satisfies the contraction condition, and denote its contraction constant by
\[
\kappa_{\mathcal{E}'}\in(0,1).
\]
Define the environment perturbation magnitude as
\[
\delta_{\mathrm{pert}}(\mathcal{E}',\mathcal{E}_{\mathrm{train}})
=
\sup_{\hat y\in\mathcal{B}}
d\!\left(
\Phi_{\mathcal{E}'}(\hat y),
\Phi_{\mathcal{E}_{\mathrm{train}}}(\hat y)
\right).
\]

Let
\[
e_t
=
d\!\left(
\hat y_t^{\mathcal{E}'},
\hat y_t^{\mathcal{E}_{\mathrm{train}}}
\right)
\]
denote the deviation between the answers at round $t$ obtained by the policy $\pi_\theta^*$ under the transferred environment and the training environment, and let $e_0=0$. By the triangle inequality,
\[
\begin{aligned}
e_t
&=
d\!\left(
\Phi_{\mathcal{E}'}(\hat y_{t-1}^{\mathcal{E}'}),
\Phi_{\mathcal{E}_{\mathrm{train}}}(\hat y_{t-1}^{\mathcal{E}_{\mathrm{train}}})
\right)\\
&\le
d\!\left(
\Phi_{\mathcal{E}'}(\hat y_{t-1}^{\mathcal{E}'}),
\Phi_{\mathcal{E}'}(\hat y_{t-1}^{\mathcal{E}_{\mathrm{train}}})
\right)
+
d\!\left(
\Phi_{\mathcal{E}'}(\hat y_{t-1}^{\mathcal{E}_{\mathrm{train}}}),
\Phi_{\mathcal{E}_{\mathrm{train}}}(\hat y_{t-1}^{\mathcal{E}_{\mathrm{train}}})
\right)\\
&\le
\kappa_{\mathcal{E}'} e_{t-1}
+
\delta_{\mathrm{pert}}.
\end{aligned}
\]
Unrolling this recurrence yields
\[
e_t
\le
\delta_{\mathrm{pert}}
\sum_{j=0}^{t-1}\kappa_{\mathcal{E}'}^j
=
\frac{\delta_{\mathrm{pert}}(1-\kappa_{\mathcal{E}'}^t)}{1-\kappa_{\mathcal{E}'}}.
\]
Therefore, the trajectory deviation between the transferred environment and the training environment is always controlled by an explicit upper bound.

Moreover, by the triangle inequality and the a priori error bound on the training environment from Proposition~1,
\[
d\!\left(
\hat y_t^{\mathcal{E}_{\mathrm{train}}},
y^*
\right)
\le
\kappa_{\mathcal{E}_{\mathrm{train}}}^t
\,d(\hat y_0,y^*),
\]
the answer error under the transferred environment satisfies
\[
\begin{aligned}
d\!\left(
\hat y_t^{\mathcal{E}'},
y^*
\right)
&\le
d\!\left(
\hat y_t^{\mathcal{E}'},
\hat y_t^{\mathcal{E}_{\mathrm{train}}}
\right)
+
d\!\left(
\hat y_t^{\mathcal{E}_{\mathrm{train}}},
y^*
\right)\\
&\le
\frac{\delta_{\mathrm{pert}}(1-\kappa_{\mathcal{E}'}^t)}{1-\kappa_{\mathcal{E}'}}
+
\kappa_{\mathcal{E}_{\mathrm{train}}}^t\,d(\hat y_0,y^*).
\end{aligned}
\]
Hence,
\[
\limsup_{t\to\infty}
d\!\left(
\hat y_t^{\mathcal{E}'},
y^*
\right)
\le
\frac{\delta_{\mathrm{pert}}}{1-\kappa_{\mathcal{E}'}}.
\]
This shows that as long as the environment perturbation $\delta_{\mathrm{pert}}$ is bounded and sufficiently small, the reasoning error after transfer remains bounded, and thus the dynamic optimization policy has stable transferability in the new environment.

If it is further assumed that the transferred environment also satisfies the correctness-preservation condition
\[
\Phi_{\mathcal{E}'}(y^*)=y^*,
\]
then $y^*$ is a fixed point of $\Phi_{\mathcal{E}'}$. Since $\Phi_{\mathcal{E}'}$ is also a contraction mapping on $\mathcal{B}$, the Banach fixed-point theorem directly gives
\[
d\!\left(
\hat y_t^{\mathcal{E}'},
y^*
\right)
\le
\kappa_{\mathcal{E}'}^t
\,d\!\left(
\hat y_0^{\mathcal{E}'},
y^*
\right).
\]
In this case, convergence in the transferred environment can be established independently, without relying on the perturbation comparison with the training environment.

Now consider the case where $\mathcal{E}'$ is stronger than $\mathcal{E}_{\mathrm{train}}$, formalized as
\[
\kappa_{\mathcal{E}'}
\le
\kappa_{\mathcal{E}_{\mathrm{train}}},
\qquad
d\!\left(
\hat y_0^{\mathcal{E}'},
y^*
\right)
\le
d\!\left(
\hat y_0^{\mathcal{E}_{\mathrm{train}}},
y^*
\right).
\]
If $\Phi_{\mathcal{E}'}(y^*)=y^*$ also holds, then
\[
d\!\left(
\hat y_t^{\mathcal{E}'},
y^*
\right)
\le
\kappa_{\mathcal{E}'}^t
d\!\left(
\hat y_0^{\mathcal{E}'},
y^*
\right)
\le
\kappa_{\mathcal{E}_{\mathrm{train}}}^t
d\!\left(
\hat y_0^{\mathcal{E}_{\mathrm{train}}},
y^*
\right).
\]
Therefore, under the above conditions, the answer-error bound in the transferred environment is no worse than the corresponding error bound in the training environment. This indicates that when the target MLLM has stronger initial estimation capability and faster error contraction, the learned dynamic optimization policy can maintain or even improve reasoning performance in the sense of error control.

Next, we analyze the stopping rounds after transfer. According to the contraction analysis in Proposition~1,
\[
d(\hat y_{t+1},\hat y_t)
\le
\kappa_{\mathcal{E}}^t\,d(\hat y_1,\hat y_0),
\]
and the stopping criterion is
\[
d(\hat y_t,\hat y_{t-1})<\varepsilon.
\]
Therefore, for any environment $\mathcal{E}$, the stopping round satisfies the upper bound
\[
T_\tau^{\mathcal{E}}
\le
\min\!\left(
1+
\left\lceil
\frac{\log(d(\hat y_1,\hat y_0)/\varepsilon)}{\log(1/\kappa_{\mathcal{E}})}
\right\rceil,
\;T
\right).
\]
If
\[
\kappa_{\mathcal{E}'}
<
\kappa_{\mathcal{E}_{\mathrm{train}}},
\]
then
\[
\log(1/\kappa_{\mathcal{E}'})
>
\log(1/\kappa_{\mathcal{E}_{\mathrm{train}}}),
\]
and thus the upper bound on the stopping rounds for the transferred environment is smaller, meaning that a stronger environment has potentially higher reasoning efficiency. Conversely, if the transferred environment is weaker, the upper bound on the stopping rounds becomes larger, but the adaptive stopping mechanism still guarantees
\[
T_\tau\le T,
\qquad
d(\hat y_{T_\tau},y^*)
\le
\frac{\kappa_{\mathcal{E}'}}{1-\kappa_{\mathcal{E}'}}\varepsilon.
\]
Therefore, regardless of whether the transferred environment is stronger or weaker, the policy can maintain controllable reasoning precision within a finite number of rounds.

In summary, environment transfer can be formalized as a bounded perturbation of the dynamic optimization operator: when the perturbation magnitude $\delta_{\mathrm{pert}}$ is sufficiently small, the post-transfer reasoning error remains bounded; when the transferred environment also satisfies contraction and correctness-preservation conditions, its convergence guarantee can be established directly; when the target environment has a smaller contraction factor and a better initial estimate, the answer-error bound after transfer is no worse than that in the training environment and admits a smaller upper bound on the stopping rounds. Therefore, the dynamic optimization policy can be effectively transferred across different MLLMs satisfying the above regularity conditions, while maintaining stable reasoning performance in the sense of error control.
\end{proof}

\section{Motivation: From Prompts to Policies}
\label{app:positioning}

\subsection{The Adaptivity Gap}
\label{app:adaptivity}
Many fixed-configuration reasoning baselines can be abstracted as selecting one intervention in a fixed space---for example, a task-level prompt, a sampling budget, or a collaboration protocol---and reusing it across task instances. Formally, let $U(p, q)$ denote the reasoning utility of applying intervention $p$ to instance $q$; such static methods solve $\sup_p \mathbb{E}_{q\sim\mathcal{D}}\,[U(p,q)]$. For any utility function and any instance distribution, it always holds that
\begin{equation}
    \sup_{p} \; \mathbb{E}_{q \sim \mathcal{D}}\left[U(p, q)\right] \;\leq\; \mathbb{E}_{q \sim \mathcal{D}}\Big[\sup_{p}\; U(p, q)\Big].
\end{equation}
The difference between the two sides characterizes the potential headroom of per-instance customized interventions over a single distribution-level intervention, in direct lineage with the adaptivity gap studied in stochastic optimization, where the benefit of adaptivity is measured between optimal adaptive and non-adaptive policies. In multimodal reasoning, instances can differ in the relevant visual evidence and required reasoning depth---for example, local regions, global structures, or cross-modal correspondences---so a single task-level intervention need not suit every instance. Under our evaluated configuration, the GEPA prompt obtained by dataset-level search reaches an overall EM of 44.47, below the 48.76 of plain inference; this illustrates the limitation of a single learned prompt in this setting. MMDynOpt-Agent instead learns a policy that maps the current state to an intervention sequence $\{\textit{dop}_t\}$ conditioned on intermediate observations. Its policy class can represent single-turn interventions as special cases while supporting multi-turn, instance-conditioned decisions. The optimization problem is therefore shifted from searching for one prompt to learning a policy over prompts, with Prism Reward incorporating budget awareness during training.

\begin{table*}[t]
\centering
\caption{Comparison of MMDynOpt-Agent with selected reasoning-enhancement paradigms.}
\label{tab:paradigm}
\vspace{-0.6em}
{\fontsize{8}{9}\selectfont
\begin{tblr}{colspec={Q[l,2.3cm]Q[l,2.4cm]Q[l,3.0cm]Q[l,2.8cm]Q[l,2.4cm]Q[l,3.4cm]},
  colsep=4pt,rowsep=1.5pt,stretch=1.0,
  row{1}={bg=white, font=\bfseries\fontsize{8}{9}\selectfont},
  hline{1,Z}={1.0pt},
  hline{2}={0.6pt}}
\textbf{Dimension} & \textbf{Post-Training} & \textbf{Automatic Prompt Opt.} & \textbf{Test-Time Scaling} & \textbf{Multi-Agent} & \textbf{MMDynOpt-Agent} \\
Representatives & SFT, GRPO & OPRO, TextGrad, GEPA & BoN, ToT, Self-Consistency & MARS & --- \\
Optimized object & Model parameters & Task-level prompt or instance-level text variable & Sampling/search allocation; some methods adapt policies or models & Prompt/protocol states; method-dependent & Policy $\pi_\theta: \mathcal{S} \to \Delta(\mathcal{A})$ \\
Conditioning granularity & --- & Task-level or instance-level via an optimization loop & Instance-level; commonly fixed, with learned variants & Instance-level; method-dependent & Instance-level $\times$ turn-level \\
Compute investment & Training time & Test-time search or optimization loop & Test-time sampling and search & Test-time multi-turn dialogue & Amortized at training; single-trajectory inference \\
Interaction depth & --- & --- & Usually a hyperparameter; learned variants exist & Usually protocol-defined; method-dependent & Action in the MDP \\
Target-model requirement & Trainable parameters & Black-box & Black-box & Black-box & Black-box; zero-shot cross-model transfer \\
\end{tblr}}
\end{table*}

\subsection{Three Separations}
\label{app:separations}
Table~\ref{tab:paradigm} contrasts MMDynOpt-Agent with selected existing paradigms along five dimensions; the following three separations characterize our framework.

\textbf{Separation 1: the output objects live in different spaces.}
Task-level prompt search methods represented by OPRO~\cite{yang2023large} and GEPA~\cite{agrawal2025gepa} output one point in the prompt space and reuse it across instances, whereas the output of our framework is a mapping $\pi_\theta: \mathcal{S} \to \Delta(\mathcal{A})$: each $\textit{dop}_t$ is generated on the fly, conditioned on the agent state $s_t$ that contains the image, the question, and the full interaction history. This distinction enables per-instance customization without an explicit prompt-search loop at inference.

\textbf{Separation 2: compute is invested at different times.}
Test-time scaling and instance-level prompt optimization can invest compute in inference-time sampling, search, or optimization loops, whereas our framework amortizes policy learning into the lightweight agent weights via GRPO. At inference, the agent performs no candidate sampling, no search, and no explicit optimization of a text variable; one sampled trajectory leads to the answer. Under the selected configurations in our efficiency comparison, our method reaches 71.77 F1 with 2.8K tokens per sample, whereas the fixed-configuration multi-step baselines consume 11K--413K tokens per sample at lower accuracy.

\textbf{Separation 3: interaction depth holds a different status.}
For fixed-configuration baselines such as Best-of-N~\cite{brown2024large}, ToT~\cite{yao2023tree}, and Self-Refine~\cite{madaan2023self}, sample count, search width, or iteration count is manually preset and shared across instances. In our framework, whether to continue steering or to terminate with an answer is itself an element of the action space, and the termination time is decided per instance by the policy. Depth therefore becomes a decision variable of the learned controller; the observed performance--cost trade-off is consistent with this design.

\subsection{Shifting the Frontier}
\label{app:frontier}
Beyond the separations above, this subsection examines the latest advances within each paradigm and clarifies how our framework relates to them.

\textbf{Automatic prompt optimization.}
The paradigm is evolving along several routes: OPRO~\cite{yang2023large} formulates prompt search as LLM-driven meta-optimization, GEPA~\cite{agrawal2025gepa} introduces genetic-Pareto evolution with reflection, TextGrad~\cite{yuksekgonul2024textgrad} iteratively optimizes prompts and instance-level solutions via textual gradients, and PromptAgent~\cite{wang2023promptagent} organizes prompt discovery from a planning perspective. These methods optimize text variables---either a task-level prompt reused across instances or an instance-level variable updated through an optimization loop. In contrast, our framework learns a policy that generates prompts: optimization occurs during training and is frozen into policy weights, while inference conditions on each instance and turn without a test-time prompt-optimization loop. This is a complementary route to instance-level prompt optimization rather than a claim that it subsumes all prompt-optimization methods.

\textbf{Test-time scaling.}
The paradigm expands test-time compute along several directions: lengthening the interaction horizon, where recent work trains agents with curriculum-based online reinforcement learning to adaptively extend interaction with the environment; widening the search structure, where step-level verifier-guided hybrid scaling combines conditional self-refinement with parallel scaling at the step level, while $\phi$-Decoding~\cite{xu2025phi}, ReST-MCTS*~\cite{zhang2024rest}, and AFLOW~\cite{zhang2024aflow} use foresight sampling, process-reward-guided tree search or self-training, and workflow search, respectively; and updating parameters at test time, where test-time reinforcement learning applies reward signals built from output frequency and entropy on unlabeled test samples to adapt the target vision-language model. These directions trade additional test-time computation for performance through longer interaction, wider search, or parameter adaptation. Our framework makes a complementary deployment choice: the target model remains frozen and inaccessible, while an external policy learns to allocate a bounded interaction budget across turns. Both interaction-horizon scaling and our framework treat interaction length as learnable, but they differ in environment, training objective, and whether the target model itself is updated. The approaches are potentially composable: under a generous budget, search or test-time adaptation could be part of the target environment, while the policy controls when to invoke further interaction.

\textbf{Multi-agent systems.}
The paradigm has evolved from division-of-labor collaboration and debate-style critique to the layered aggregation of MoA~\cite{wang2024mixture}, the cooperative reinforcement-learning fine-tuning of CORY~\cite{ma2024coevolving}, and the Socratic prompt optimization of MARS~\cite{zhang2025mars}, while MacNet~\cite{qian2024scaling} reports logistic collaborative scaling as the number of agents grows. The dividing line from our framework lies in the optimized object and learning signal. MARS uses a multi-agent prompt-optimization loop with planner and teacher--critic--student feedback, whereas our method learns one external policy that generates interventions for a frozen target model. CORY trains collaborating copies of the target LLM and therefore requires parameter access. Our framework instead assigns decisions to a lightweight policy and treats the target model as a black-box environment, which is compatible with zero-shot transfer across the evaluated target models.

\subsection{Four Hypotheses, Supporting Evidence}
\label{app:attribution}
A framework that combines reinforcement learning, multi-turn interaction, and reward design naturally raises an attribution question: do the gains come from the learned policy or from individual components? The following analyses provide supporting evidence while not establishing a complete causal decomposition.

\textbf{Hypothesis 1: the gains come from the prompt template rather than from learning.}
Under the w/o R.L. ablation, OmniMed performance decreases from 84.64 to 8.59 EM. This supports the importance of reinforcement-learning-based policy training under the evaluated protocol.

\textbf{Hypothesis 2: multi-turn interaction alone suffices and no training signal is needed.}
The comparison with the training-free MARS baseline tests whether the selected multi-turn protocol alone reproduces the observed gains. Under our implementation and evaluation settings, it does not match the performance of the learned policy. This comparison is evidence for the value of learned policy optimization in this setting, rather than a general claim about all multi-agent methods.

\textbf{Hypothesis 3: the gains come from a larger inference budget.}
Under the selected configurations, our method uses 2.8K tokens per sample and attains the highest average accuracy among the compared multi-step baselines. This is consistent with improved inference-time efficiency, while a budget-matched study is required to isolate the effect of budget allocation.

\textbf{Hypothesis 4: the policy memorizes prompt patterns specific to the training-time target model.}
If the policy relied only on patterns specific to the training-time target model, its gains would be expected to diminish after replacement. Our transfer experiments show zero-shot transfer to four target MLLMs unseen during training---InternVL3.5, Qwen3-VL-30B-A3B, Gemma-3, and Nemotron---with improvements across the evaluated model families and scales. This provides evidence against strictly target-model-specific prompting, while not by itself excluding task- or dataset-specific effects.

Together, these analyses support the interpretation that the learned, budget-aware, per-instance dynamic optimization policy is a material contributor to the observed gains. They do not replace controlled budget-matching or broader causal analyses.

\section{Prompt Used in MMDynOpt-Agent}
\label{app:prompts}

\begin{figure*}[t]
  \centering
  \includegraphics[width=1\linewidth]{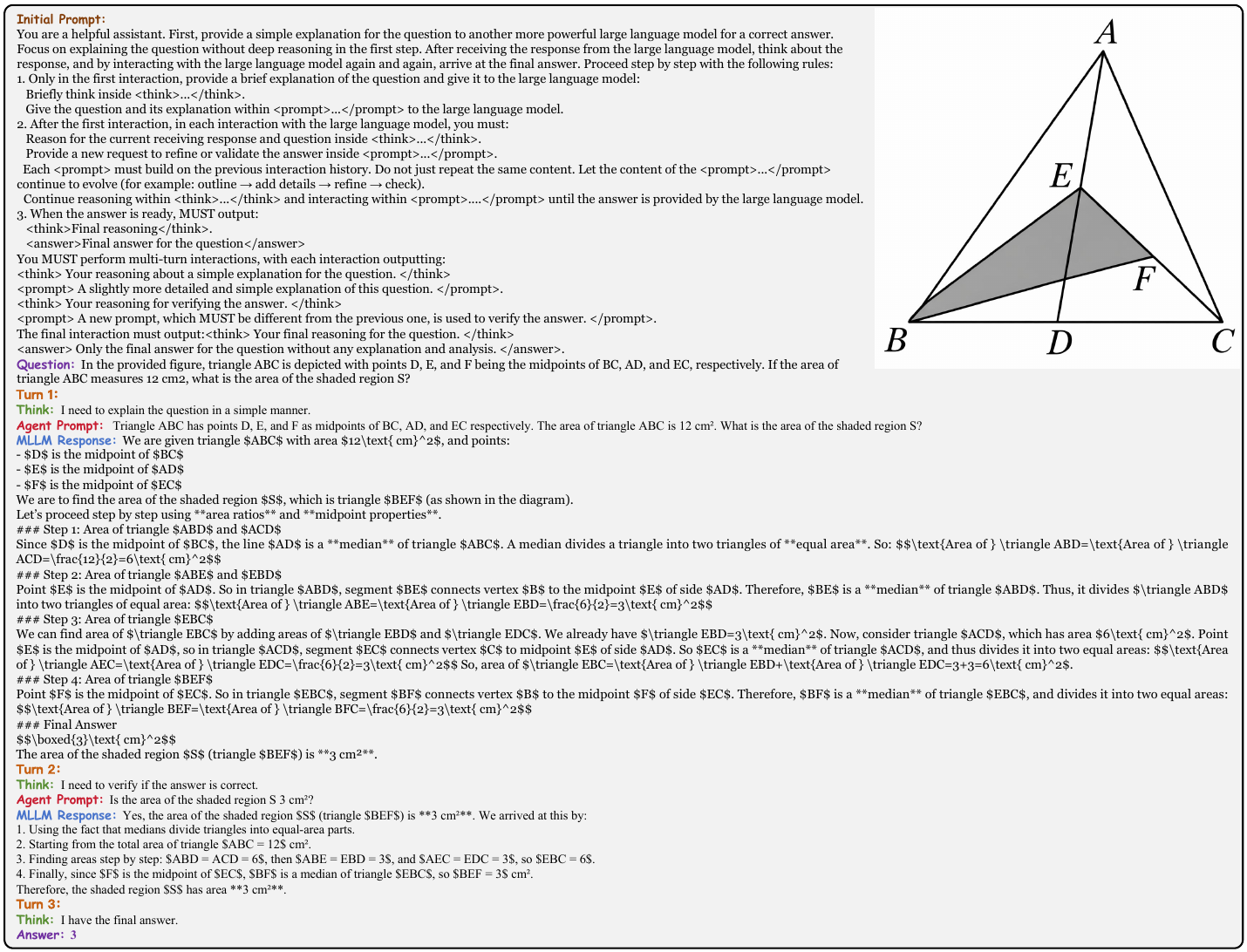}
  \vspace{-5mm}
  \caption{Workflow of MMDynOpt-Agent for enhancing MLLM multimodal reasoning through dynamic optimization prompts.}
    \label{fig: workflow}
\end{figure*}

The initial prompt template is shown in Table \ref{tab:sys-prompt-tem}. MMDynOpt-Agent improves the reasoning performance of MLLMs by continuously interacting with the MLLM to generate dynamic optimization prompts. As illustrated in Figure \ref{fig: workflow}, we provide the complete workflow of a case in which MMDynOpt-Agent enhances the multimodal reasoning performance of the target MLLM.

\section{Dataset Details}
\label{app:datasets}
We evaluate MMDynOpt-Agent on 15 multimodal reasoning benchmarks covering medicine, finance, and general science domains. Among them, 9 datasets are used for training, and 6 are reserved exclusively for out-of-distribution (OOD) generalization evaluation.

The training datasets include:
\begin{itemize}[leftmargin=*]
\item \textbf{PMC-VQA} \cite{zhang2023pmc}: A large-scale medical visual question answering dataset covering various imaging modalities and diseases.
\item \textbf{MedXpertQA} \cite{zuo2025medxpertqa}: A highly challenging benchmark for evaluating expert-level medical knowledge and advanced reasoning, spanning multiple specialties and body systems.
\item \textbf{OmniMedVQA} \cite{hu2024omnimedvqa}: A large-scale comprehensive evaluation benchmark for medical large vision-language models, covering diverse medical imaging modalities and human body regions.
\item \textbf{FinChart-Bench} \cite{shu2025finchart}: A benchmark for evaluating the financial chart comprehension capabilities of vision-language models across diverse chart types and analytical tasks.
\item \textbf{FinMME} \cite{luo2025finmme}: A benchmark dataset for financial multi-modal reasoning evaluation, covering multi-modal understanding and reasoning tasks across diverse financial scenarios.
\item \textbf{Sujet-Finance-QA-Vision-100k} \cite{Sujet-Finance-QA-Vision-100k}: A large-scale dataset for financial document visual question answering, derived from financial document images.
\item \textbf{GeoQA} \cite{chen2021geoqa}: A geometric question answering benchmark toward multimodal numerical reasoning, requiring models to parse geometric diagrams and perform mathematical calculations.
\item \textbf{ScienceQA} \cite{lu2022learn}: A multimodal science question answering dataset with annotated thought chains for learning to explain, covering multiple natural science subjects.
\item \textbf{Multi} \cite{zhu2025multi}: A multimodal understanding leaderboard with text and images, evaluating model comprehension of composite multimodal information across diverse topics.
\end{itemize}

The out-of-distribution evaluation datasets include:
\begin{itemize}[leftmargin=*]
\item \textbf{MathVista} \cite{lu2023mathvista}: A benchmark for evaluating mathematical reasoning of foundation models in visual contexts, encompassing geometry, statistics, algebra, and other mathematical tasks.
\item \textbf{PathVQA} \cite{he2020pathvqa}: A pathology visual question answering dataset constructed from pathology textbooks and digital libraries.
\item \textbf{SeePhys} \cite{xiang2025seephys}: A vision-based physics reasoning benchmark that investigates whether visual perception assists physical understanding and reasoning.
\item \textbf{VisFinEval} \cite{liu2025visfineval}: A scenario-driven Chinese multimodal benchmark for holistic financial understanding, covering diverse real-world financial visual comprehension tasks.
\item \textbf{AI2D} \cite{kembhavi2016diagram}: A diagram understanding dataset containing grade-school science diagrams with corresponding questions that require interpreting visual scientific representations.
\item \textbf{VQA-RAD} \cite{lau2018dataset}: A clinically generated dataset of visual questions and answers about radiology images, covering head, chest, and abdomen.
\end{itemize}

To ensure consistency across datasets and maintain manageable training and evaluation workloads, we randomly sample 1,280 instances from each training dataset for training (9 datasets, 11,520 instances in total) and randomly sample 128 instances from each dataset for evaluation (15 datasets, 1,920 instances in total).

\section{Baseline Details}
\label{app:baselines}

Our experiments compare MMDynOpt-Agent with two groups of baselines using different model configurations, which are as follows.

\subsection{Baselines with Qwen2.5-VL-3B}

\begin{itemize}[leftmargin=*]
\item \textbf{Baseline}: Direct inference using Qwen2.5-VL-3B-Instruct on multimodal inputs without any post-training or prompt optimization, serving as a lower bound baseline for the small model.
\item \textbf{SFT} \cite{zhang2026instruction}: Supervised fine-tuning of Qwen2.5-VL-3B on the training data using question-answer pairs to directly optimize model parameters.
\item \textbf{GRPO} \cite{guo2025deepseek}: Group Relative Policy Optimization on Qwen2.5-VL-3B using the training data, optimizing reasoning ability through reinforcement learning signals.
\end{itemize}

\definecolor{grpA}{rgb}{0.9412,0.9804,0.9490}
\definecolor{grpB}{rgb}{0.9608,0.9765,0.9980}
\definecolor{grpC}{rgb}{0.985, 0.97, 0.995}
\begin{table*}[t]
\centering
\caption{Hyperparameter configurations for MMDynOpt-Agent and the baselines in our experiments. -- indicates not applicable.}
\label{tab:hyperparams}
\vspace{-0.6em}
\begin{tblr}{colspec={Q[l,3.2cm]Q[c,4.8cm]Q[c,1.8cm]Q[c,2.0cm]Q[c,3cm]Q[c,1.2cm]},
  colsep=4pt,rowsep=1.5pt,stretch=1.0,
  row{1}={bg=white, font=\bfseries\fontsize{8}{9}\selectfont},
  row{2-4}={font=\fontsize{8}{9}\selectfont},
  row{5-13}={font=\fontsize{8}{9}\selectfont},
  row{14}={font=\fontsize{8}{9}\selectfont, bg=grpC},
  hline{1,Z}={1.0pt},
  hline{2}={0.6pt},
  hline{5}={0.4pt},
  hline{14}={0.4pt}}
\textbf{Method} & \textbf{Backbone} & \textbf{Batch Size} & \textbf{Max Length} & \textbf{Top-K} & \textbf{Epochs} \\
Baseline          & Qwen2.5-VL-3B-Instruct  & --           & 8192 & N/A                     & -- \\
SFT               & Qwen2.5-VL-3B-Instruct  & 128          & 8192 & N/A                     & 2  \\
GRPO              & Qwen2.5-VL-3B-Instruct  & 128          & 8192 & N/A                     & 1  \\
Baseline          & Qwen3-VL-32B-Instruct   & --           & 8192 & N/A                     & -- \\
CoT               & Qwen3-VL-32B-Instruct   & --           & 8192 & N/A                     & -- \\
OPRO              & Qwen3-VL-32B-Instruct   & --           & 8192 & 20 steps $\times$ 8 instr & -- \\
GEPA              & Qwen3-VL-32B-Instruct   & --           & 8192 & N/A                     & -- \\
BON               & Qwen3-VL-32B-Instruct   & --           & 8192 & $N$\,=\,10              & -- \\
S-CON             & Qwen3-VL-32B-Instruct   & --           & 8192 & $n$\,=\,10              & -- \\
S-COR             & Qwen3-VL-32B-Instruct   & --           & 8192 & 10 iterations            & -- \\
ToT               & Qwen3-VL-32B-Instruct   & --           & 8192 & 3 steps / 5 branch / 3 beam & -- \\
MARS              & Qwen3-VL-32B-Instruct   & --           & 8192 & 10 iterations            & -- \\
MMDynOpt-Agent (ours) & \makecell[c]{Agent: Qwen2.5-VL-3B-Instruct\\MLLM: Qwen3-VL-32B-Instruct} & 128 & 8192 & 5\,/\,Turn & 1  \\
\end{tblr}
\end{table*}

\subsection{Baselines with Qwen3-VL-32B}

\begin{itemize}[leftmargin=*]
\item \textbf{Baseline}: Direct inference using Qwen3-VL-32B-Instruct on multimodal inputs without any prompt optimization or multi-turn interaction, serving as a lower bound baseline for the target MLLM.
\item \textbf{CoT} \cite{mitra2024compositional}: Compositional Chain-of-Thought prompting that guides MLLMs to perform step-by-step reasoning through structured prompting templates.
\item \textbf{OPRO} \cite{yang2023large}: Optimization by PROmpting that formalizes prompt optimization as an LLM-driven search process over the prompt space.
\item \textbf{GEPA} \cite{agrawal2025gepa}: Genetic-Pareto reflective prompt evolution that automatically optimizes prompts via evolutionary algorithms and reflection mechanisms.
\item \textbf{Best-of-N (BON)} \cite{brown2024large}: Sampling $N$ candidate answers per question and selecting the best answer as the final output.
\item \textbf{Self-Consistency (S-CON)} \cite{wang2022self}: Improving the consistency and reliability of chain-of-thought reasoning through multiple sampling and majority voting.
\item \textbf{Self-Correct (S-COR)} \cite{welleck2022generating}: Iteratively improving reasoning outputs through self-feedback generation-revision cycles.
\item \textbf{Tree of Thoughts (ToT)} \cite{yao2023tree}: A tree-search-based reasoning method that enables deliberate problem solving through step-level exploration and backtracking.
\item \textbf{MARS} \cite{zhang2025mars}: Multi-Agent Adaptive Reasoning with Socratic guidance that performs automated prompt optimization through multi-agent collaboration.
\end{itemize}

\section{Evaluation Metrics}
\label{app:metrics}

We evaluate MMDynOpt-Agent and the baselines by four metrics:

\noindent\textbf{(i) Exact Match (EM).} EM measures whether the predicted answer exactly matches the ground truth. Let $\text{norm}(\cdot)$ denote the normalization function:
\begin{equation}
\text{EM} = \frac{1}{N} \sum_{i=1}^{N} \mathbb{1}\{\text{norm}(\hat{y}_i) = \text{norm}(y_i^*)\}.
\end{equation}

\noindent\textbf{(ii) F1 Score (F1).} The F1 score measures the token-level overlap between the predicted answer $\hat{y}_i$ and the ground-truth answer $y_i^*$ using the harmonic mean of precision and recall:
\begin{equation}
\text{F1} = \frac{1}{N} \sum_{i=1}^{N} \frac{2 \cdot |\text{tokens}(\hat{y}_i) \cap \text{tokens}(y_i^*)|}{|\text{tokens}(\hat{y}_i)| + |\text{tokens}(y_i^*)|}.
\end{equation}

\noindent\textbf{(iii) Semantic Similarity (SSim).} SSim assesses the semantic similarity between the predicted answer and the ground-truth answer. Let $\text{Enc}(\cdot)$ be the semantic embedding function:
\begin{equation}
\text{SSim} = \frac{1}{N} \sum_{i=1}^{N} \cos\!\left(\text{Enc}(\hat{y}_i),\; \text{Enc}(y_i^*)\right).
\end{equation}

\noindent\textbf{(iv) Total Tokens per F1 (FPT).} FPT measures the performance-cost trade-off efficiency, defined as the average total tokens consumed per unit of F1 score gained:
\begin{equation}
\text{FPT} = \frac{\text{Total Tokens per Sample}}{\text{F1 Score}},
\end{equation}
where the total tokens include both input tokens sent to and output tokens received from the target MLLM. A lower FPT indicates higher reasoning efficiency.

\begin{figure*}[t]
  \centering
  \includegraphics[width=0.88\linewidth]{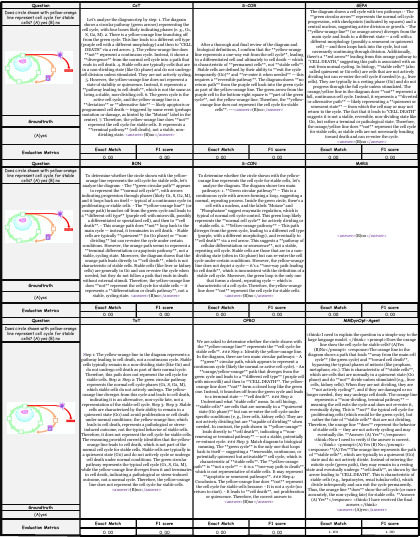}
  \caption{Case studies of multimodal reasoning methods, including CoT, automatic prompt optimization methods (OPRO, GEPA), test-time scaling methods (BoN, S-CoT, S-COR, ToT), MAS (MARS), and our proposed framework MMDynOpt-Agent.}
    \label{fig: case_study}
\end{figure*}

\section{Implementation Details}
\label{app:implementation}
As shown in Table~\ref{tab:hyperparams}, we summarize the detailed hyperparameter configurations used throughout our experiments, including the model backbone, batch size, input limits, test-time scaling budget, and training setup. All hyperparameters are set appropriately. For the Prism Reward, we set the format reward components $\alpha_{\text{mid}}=0.3$, $\alpha_{\text{tag}}=0.15$, $\alpha_{\text{ne}}=0.15$, and $\alpha_{\text{full}}=0.1$, the budget efficiency weight $\lambda_b=0.1$, the answer quality threshold $\eta=0.8$, and the normalization bounds $N_{\text{max}}=5$, $L_{\text{in}}^{\max}=500$, and $L_{\text{out}}^{\max}=1000$ tokens, with the maximum number of dynamic optimization rounds $T=5$ and the group size $G=4$ during training. For the main comparison experiments, we conduct three independent runs and report the mean and standard deviation of the results.

\section{MMDynOpt-Agent Algorithm Details}
\label{app:algorithm_details}

To illustrate the mechanism of MMDynOpt-Agent, we present its full workflow in Algorithm~\ref{alg:mmdynopt}, comprising three key phases:
\textbf{(1) Environment and Agent Initialization.}
The target MLLM $\mathcal{E}$ serves as the environment, whose parameters and gradients are inaccessible to the agent $\pi_\theta$. The agent constructs the initial state $s_0$ from the multimodal input $q=(q_{\text{text}}, q_{\text{img}})$ and the system instruction.
\textbf{(2) Multi-turn Dynamic Optimization Reasoning.}
Given input $q$, the agent adaptively selects between a steering action and an answering action over up to $T$ rounds: a steering action sends a dynamic optimization prompt $\textit{dop}_t$ to $\mathcal{E}$ and receives a reasoning response $o_t$; an answering action outputs the final answer and terminates the interaction.
\textbf{(3) End-to-end Policy Optimization with Prism Reward.}
The policy is optimized using GRPO, where Prism Reward jointly constrains format compliance, answer correctness, and budget efficiency via hierarchical gating. The objective $\mathcal{J}_{\text{GRPO}}$ is computed from clipped advantage-weighted updates over sampled trajectories.

\begin{algorithm}[t]
\caption{MMDynOpt-Agent: Dynamic Optimization via RL}
\label{alg:mmdynopt}
\small
\begin{flushleft}
\textbf{Require:} Multimodal input $q=(q_{\text{text}}, q_{\text{img}})$, target MLLM $\mathcal{E}$, agent policy $\pi_\theta$, reward function $R(\tau)$\\
\textbf{Ensure:} Final answer $\hat{y}$
\end{flushleft}
\vspace{2pt}
\begin{flushleft}
\hspace{0.5em}\textbf{// 1: Environment and Agent Initialization}\\
\hspace{0.5em}1: Initialize environment interaction history $h_0^{\mathcal{E}} \leftarrow \emptyset$\\
\hspace{0.5em}2: Construct initial state $s_0 \leftarrow \text{Encode}\!\left([\text{sys};\; q_{\text{text}};\; q_{\text{img}}]\right)$\\
\hspace{0.5em}3: Initialize trajectory $\tau \leftarrow \emptyset$\\[3pt]
\hspace{0.5em}\textbf{// 2: Multi-turn Dynamic Optimization Reasoning}\\
\hspace{0.5em}4: \textbf{for} $t = 0$ to $T-1$ \textbf{do}\\
\hspace{0.5em}5: \hspace{1.5em}Generate reasoning chain: $\text{think}_t \sim \pi_\theta(\cdot \mid s_t)$\\
\hspace{0.5em}6: \hspace{1.5em}Choose action intent: $\alpha_t \sim \pi_\theta(\cdot \mid \text{think}_t, s_t)$\\
\hspace{0.5em}7: \hspace{1.5em}\textbf{if} $\alpha_t = (\texttt{answer})$ \textbf{then}\\
\hspace{0.5em}8: \hspace{3.0em}Output answer: $\hat{y} \sim \pi_\theta(\cdot \mid \text{think}_t, s_t)$\\
\hspace{0.5em}9: \hspace{3.0em}$\tau \leftarrow \tau \cup \{(s_t, \text{think}_t, \hat{y})\}$; \;\textbf{return} $\hat{y}$\\
10: \hspace{1.5em}\textbf{else if} $\alpha_t = (\texttt{steering})$ \textbf{then}\\
11: \hspace{3.0em}Generate dynamic optimization prompt: $\textit{dop}_t \sim \pi_\theta(\cdot \mid \text{think}_t, s_t)$\\
12: \hspace{3.0em}Environment reasoning: $o_t \leftarrow \mathcal{E}\!\left(\textit{dop}_t,\; q_{\text{img}} \cdot \mathbbm{1}[t{=}0],\; h_t^{\mathcal{E}}\right)$\\
13: \hspace{3.0em}Update environment history: $h_{t+1}^{\mathcal{E}} \leftarrow h_t^{\mathcal{E}} \cup \{(\textit{dop}_t, o_t)\}$\\
14: \hspace{3.0em}Update state: $s_{t+1} \leftarrow [s_t;\; \text{think}_t;\; \textit{dop}_t;\; o_t]$\\
15: \hspace{3.0em}$\tau \leftarrow \tau \cup \{(s_t, \text{think}_t, \textit{dop}_t, o_t)\}$\\
16: \hspace{1.5em}\textbf{end if}\\
17: \hspace{0em}\textbf{end for}\\
18: Output forced terminal answer: $\hat{y} \sim \pi_\theta(\cdot \mid s_T)$; \;\textbf{return} $\hat{y}$\\[3pt]
\hspace{0.5em}\textbf{// 3: End-to-end Policy Optimization (GRPO)}\\
19: Sample $G$ trajectories $\{\tau_i\} \sim \pi_{\theta_{\text{old}}}$\\
20: \textbf{for each} $\tau_i$ \textbf{do}\\
21: \hspace{1.5em}Compute Prism Reward: $R(\tau_i) = -1 + R_{\text{format}}(\tau_i) + \mathbbm{1}\{R_{\text{format}} = 1\} \cdot R_{\text{answer}}(\hat{y}, y^*)$\\
\hspace{3.0em}where $R_{\text{answer}} = r_a + \lambda_b \cdot r_a \cdot r_b \cdot \mathbbm{1}[r_a > \eta]$\\
22: \hspace{1.5em}Compute advantage: $\hat{A}(\tau_i) = \frac{R(\tau_i) - \text{mean}(\{R(\tau_j)\})}{\text{std}(\{R(\tau_j)\})}$\\
23: \textbf{end for}\\
24: Update policy via GRPO: $\mathcal{J}_{\text{GRPO}} \sim \sum_{i=1}^{G} \sum_{t=1}^{|\tau_i|} m_t^{(i)} \cdot \min\!\left(\rho_\theta(a_t^{(i)}) \hat{A}(\tau_i),\; \text{clip}(\rho_\theta(a_t^{(i)}), 1 \pm \epsilon) \hat{A}(\tau_i)\right)$\\
25: where $\rho_\theta(a_t^{(i)}) = \frac{\pi_\theta(a_t^{(i)} \mid s_{t-1}^{(i)})}{\pi_{\theta_{\text{old}}}(a_t^{(i)} \mid s_{t-1}^{(i)})}$, \; $m_t^{(i)} \in \{0, 1\}$ is the token attribution mask ($m_t^{(i)}=1$ iff token $t$ is from $\pi_\theta$)
\end{flushleft}
\end{algorithm}

\textbf{Complexity Analysis.}
MMDynOpt-Agent involves three computational components corresponding to the above phases.
First, the initialization phase encodes the multimodal input with complexity $O(L_0)$, where $L_0$ is the token length of the initial state.
Second, during multi-turn dynamic optimization reasoning, the agent performs up to $T$ rounds of interaction, each involving autoregressive generation by the agent and a reasoning call to the target MLLM; since the agent is a lightweight model (e.g., Qwen2.5-VL-3B), its inference cost is substantially lower than that of the target MLLM, and the adaptive stopping mechanism ensures the actual number of rounds $T_\tau \leq T$ adjusts to problem difficulty.
Third, for policy optimization, GRPO processes $G$ sampled trajectories of maximum length $T$, with gradient update cost $O(G \cdot T \cdot |\theta|)$, where $|\theta|$ denotes the number of agent parameters. Since gradients propagate only through agent-generated tokens via the token attribution mask $m_t^{(i)}$, excluding observation tokens from $\mathcal{E}$, the effective per-step gradient computation is reduced.
Each component is computationally tractable and benefits from batch sampling and parallel inference.

\section{Case Study}
\label{app:case_study}
As shown in Figure \ref{fig: case_study}, existing baselines fail to provide the correct answer because they misinterpret the yellow-orange pathway as a terminal death trajectory rather than the functional state of stable cells. Baseline, Chain-of-Thought, S-CoT, GEPA, BoN, ToT, and OPRO all fall short, indicating limited grounding and semantic alignment despite enhanced reasoning or optimization strategies. In contrast, MMDynOpt-Agent correctly captures the semantic role of the pathway and yields the right answer, showing that dynamic prompt optimization is more effective for multimodal reasoning.

\section{Budget-Matched Comparison}
\label{app:budget}

To examine whether the performance advantage stems from budget allocation rather than from the learned policy, this section sweeps the sampling budgets of BoN and Self-Consistency from low to high and measures the accuracy and cost at each budget level in Table~\ref{tab:budget}, under the same target-MLLM-side accounting as Table~\ref{tab:efficiency}.

\begin{table}[t]
\centering
\caption{Accuracy and per-sample token cost of BoN and Self-Consistency under increasing sampling budgets.}
\label{tab:budget}
\vspace{-0.6em}
{\fontsize{8}{9}\selectfont
\begin{tblr}{colspec={Q[l,1.75cm]Q[c,0.42cm]Q[c,0.7cm]Q[c,0.7cm]Q[c,0.95cm]Q[c,0.9cm]Q[c,1.0cm]},
  colsep=3pt,rowsep=1pt,stretch=1.0,
  row{1}={bg=white, font=\bfseries\fontsize{8}{9}\selectfont},
  row{2-9}={font=\fontsize{8}{9}\selectfont},
  row{10}={font=\bfseries\fontsize{8}{9}\selectfont, bg=grpC},
  hline{1,Z}={1.0pt},
  hline{2}={0.6pt}}
\textbf{Method} & \textbf{N} & \textbf{EM} & \textbf{F1} & \textbf{Input} & \textbf{Output} & \textbf{Total} \\
BON & 2 & 50.21 & 64.73 & 2052.2 & 1389.8 & 3442.0 \\
BON & 4 & 49.32 & 63.76 & 4105.3 & 2885.9 & 6991.2 \\
BON & 8 & 49.22 & 63.95 & 8210.0 & 5598.5 & 13808.6 \\
BON & 10 & 49.97 & 64.10 & 10260.3 & 6756.8 & 17017.1 \\
S-CON & 2 & 49.74 & 64.12 & 896.5 & 1379.7 & 2276.2 \\
S-CON & 4 & 51.04 & 65.30 & 1792.9 & 2752.8 & 4545.7 \\
S-CON & 8 & 51.35 & 65.52 & 3585.0 & 5400.3 & 8985.4 \\
S-CON & 10 & 52.08 & 66.05 & 4479.5 & 6644.5 & 11124.0 \\
Ours & -- & 59.48 & 71.77 & 1911.6 & 861.5 & 2773.1 \\
\end{tblr}}
\end{table}

\textbf{Matching the budget does not close the gap.}
The table contains a direct comparison at comparable cost: at its lowest budget level, Self-Consistency consumes slightly fewer tokens per sample than our method, yet its accuracy falls far below ours. Moreover, our operating point is bracketed by the two lowest Self-Consistency budgets, so the accuracy interpolated at our exact expenditure is bounded by theirs and remains well below ours. Aligning the budget does not close the gap. At comparable expenditure, a clear accuracy margin remains between the learned policy and fixed-configuration sampling.

\textbf{Additional budget yields limited returns.}
As the number of samples grows, the cost of both baseline curves multiplies while accuracy improves only marginally: the marginal gains of Self-Consistency are limited overall and quickly plateau, while BoN does not improve with additional budget. Both curves remain well below the accuracy of our method, and within the tested range their growth trends show no sign of approaching our operating point.

\textbf{Attribution.}
The gap persists when the budget is aligned, and persists after the budget is multiplied; the growth trends of the curves give no indication that further scaling would close it. The performance advantage is therefore unlikely to stem from budget allocation and aligns instead with how the budget is used: fixed-configuration sampling spreads its expenditure evenly over mutually independent retries, whereas the learned policy invests it in targeted interaction conditioned on intermediate feedback. Under the reported token accounting, this comparison provides evidence against a purely resource-based explanation of the advantage.

\section{Visual Contribution Analysis}
\label{app:visual}

To quantify the contribution of visual information to the framework, this section removes images from the entire pipeline. Both the agent and the target model receive only the question text. We re-evaluate on the fifteen datasets and contrast the results with those under full multimodal input in Table~\ref{tab:visual}. This setting corresponds to the text-only evaluation protocol used in visual question answering research to probe language priors.

\begin{table}[t]
\centering
\caption{Per-dataset accuracy with and without image input.}
\label{tab:visual}
\vspace{-0.6em}
{\fontsize{8}{9}\selectfont
\begin{tblr}{colspec={Q[l,1.8cm]Q[c,0.62cm]Q[c,0.62cm]Q[c,0.62cm]Q[c,0.62cm]Q[c,0.8cm]Q[c,0.8cm]},
  colsep=3pt,rowsep=1pt,stretch=1.0,
  row{1}={bg=white, font=\bfseries\fontsize{8}{9}\selectfont},
  row{2}={bg=white, font=\bfseries\fontsize{8}{9}\selectfont},
  row{3-17}={font=\fontsize{8}{9}\selectfont},
  row{18}={font=\bfseries\fontsize{8}{9}\selectfont, bg=grpC},
  cell{1}{2}={c=2}{halign=c},
  cell{1}{4}={c=2}{halign=c},
  cell{1}{6}={r=2}{halign=c},
  cell{1}{7}={r=2}{halign=c},
  hline{1,Z}={1.0pt},
  hline{3}={0.6pt}}
\textbf{Dataset} & \textbf{w/ Image} & & \textbf{w/o Image} & & \textbf{$\Delta$EM} & \textbf{$\Delta$F1} \\
 & \textbf{EM} & \textbf{F1} & \textbf{EM} & \textbf{F1} & & \\
FinChart-Bench & 57.81 & 72.50 & 3.12 & 7.71 & $-$54.69 & $-$64.79 \\
FinMME & 51.56 & 65.61 & 2.34 & 5.63 & $-$49.22 & $-$59.98 \\
GeoQA & 39.58 & 68.52 & 28.91 & 53.72 & $-$10.67 & $-$14.80 \\
Multi & 66.15 & 72.24 & 59.38 & 64.51 & $-$6.77 & $-$7.73 \\
MathVista & 70.31 & 75.39 & 19.53 & 24.42 & $-$50.78 & $-$50.97 \\
MedXpertQA & 43.75 & 52.73 & 30.47 & 40.59 & $-$13.28 & $-$12.14 \\
OmniMedVQA & 84.64 & 88.77 & 42.97 & 48.25 & $-$41.67 & $-$40.52 \\
PMC-VQA & 78.12 & 82.93 & 34.38 & 44.24 & $-$43.74 & $-$38.69 \\
PathVQA & 81.51 & 81.51 & 66.41 & 66.41 & $-$15.10 & $-$15.10 \\
ScienceQA & 96.61 & 97.77 & 71.88 & 78.28 & $-$24.73 & $-$19.49 \\
SeePhys & 3.91 & 48.87 & 0.78 & 31.88 & $-$3.13 & $-$16.99 \\
VisFinEval & 79.17 & 81.03 & 37.50 & 43.88 & $-$41.67 & $-$37.15 \\
AI2D & 92.19 & 94.06 & 67.97 & 73.14 & $-$24.22 & $-$20.92 \\
FinQA & 4.69 & 39.83 & 0.00 & 24.01 & $-$4.69 & $-$15.82 \\
VQA-RAD & 42.19 & 54.73 & 14.84 & 19.69 & $-$27.35 & $-$35.04 \\
All & 59.48 & 71.77 & 32.03 & 41.76 & $-$27.45 & $-$30.01 \\
\end{tblr}}
\end{table}

\textbf{Visual information carries a substantial share of performance.}
After removing images, both overall metrics drop sharply, and the decline is universal: EM and F1 fall in the same direction on every one of the fifteen datasets without exception. The performance of the framework is built on substantive use of the visual input beyond language priors or benchmark-format patterns. If the gains came mainly from text-side shortcuts, removing images would not cause such a widespread and deep collapse.

\textbf{The degradation aligns with the irreplaceability of visual evidence.}
The per-dataset degradation exhibits a clear gradient. Tasks whose evidence exists only in images collapse most deeply: chart understanding nearly fails outright, while MathVista, OmniMedVQA, PMC-VQA, VQA-RAD, and VisFinEval lose roughly half or more of their F1. Tasks whose question text itself carries substantial information degrade mildly. The questions of knowledge-oriented medical QA carry extensive clinical descriptions in text, and the problem statements of geometry questions restate most of the geometric conditions, so both drop far less than the perception-intensive tasks. The gradient matches the degree to which the required evidence is exclusive to the image, indicating that the gains of the framework concentrate where visual information matters most.

\section{Interaction Depth Analysis}
\label{app:depth}

Based on evaluation trajectories collected across the fifteen datasets, this section examines the per-dataset distribution of interaction rounds in Table~\ref{tab:depth}, the relation between depth allocation and instance difficulty, and the single failure case.

\begin{table}[t]
\centering
\caption{Per-dataset distribution of interaction rounds. Escalation denotes trajectories that take more than two rounds.}
\label{tab:depth}
\vspace{-0.6em}
{\fontsize{8}{9}\selectfont
\begin{tblr}{colspec={Q[l,1.9cm]Q[c,0.68cm]Q[c,0.68cm]Q[c,0.58cm]Q[c,0.58cm]Q[c,0.68cm]Q[c,0.95cm]},
  colsep=3pt,rowsep=1pt,stretch=1.0,
  row{1}={bg=white, font=\bfseries\fontsize{8}{9}\selectfont},
  row{2-16}={font=\fontsize{8}{9}\selectfont},
  row{17}={font=\bfseries\fontsize{8}{9}\selectfont, bg=grpC},
  hline{1,Z}={1.0pt},
  hline{2}={0.6pt}}
\textbf{Dataset} & \textbf{Mean} & \textbf{Std} & \textbf{Min} & \textbf{Max} & \textbf{Esc.} & \textbf{Ratio} \\
FinChart-Bench & 2.00 & 0.00 & 2 & 2 & 0 & 0.0\% \\
FinMME & 2.00 & 0.00 & 2 & 2 & 0 & 0.0\% \\
GeoQA & 2.05 & 0.23 & 2 & 3 & 7 & 5.5\% \\
Multi & 2.05 & 0.30 & 2 & 5 & 4 & 3.1\% \\
MathVista & 2.02 & 0.20 & 2 & 4 & 2 & 1.6\% \\
MedXpertQA & 2.01 & 0.09 & 2 & 3 & 1 & 0.8\% \\
OmniMedVQA & 2.01 & 0.09 & 2 & 3 & 1 & 0.8\% \\
PMC-VQA & 2.00 & 0.00 & 2 & 2 & 0 & 0.0\% \\
PathVQA & 2.00 & 0.00 & 2 & 2 & 0 & 0.0\% \\
ScienceQA & 1.98 & 0.18 & 0 & 2 & 0 & 0.0\% \\
SeePhys & 2.07 & 0.28 & 2 & 4 & 8 & 6.2\% \\
VisFinEval & 2.01 & 0.09 & 2 & 3 & 1 & 0.8\% \\
AI2D & 2.01 & 0.09 & 2 & 3 & 1 & 0.8\% \\
FinQA & 2.01 & 0.09 & 2 & 3 & 1 & 0.8\% \\
VQA-RAD & 2.00 & 0.00 & 2 & 2 & 0 & 0.0\% \\
All & 2.01 & 0.15 & 0 & 5 & 26 & 1.4\% \\
\end{tblr}}
\end{table}

\textbf{Depth is short and stable.}
Across all trajectories, the number of interaction rounds averages 2.01 with a standard deviation of 0.15, and on five datasets every trajectory takes exactly two rounds. The learned policy converges to a short and highly stable interaction depth, consistent with the training-time convergence of model calls to about two, and directly underlying the low inference cost of 2.8K tokens per sample. Concentrating depth at such a short value is consistent with the behavior that a budget-aware objective is designed to elicit: minimizing interaction overhead while preserving answer quality.

\textbf{Escalation concentrates on multi-step reasoning benchmarks.}
Trajectories exceeding two rounds account for only 1.4\% overall, and their distribution is highly uneven. Escalation concentrates on benchmarks of mathematical, geometric, physical, and composite reasoning: 6.2\% on SeePhys, 5.5\% on GeoQA, 3.1\% on Multi, and 1.6\% on MathVista, with the deepest trajectories reaching four to five rounds. The escalation ratio is at most 0.8\% on the remaining datasets and exactly zero on six of them. The policy completes the vast majority of instances in two rounds and spends extra rounds predominantly on multi-step reasoning instances. This on-demand allocation of interaction rounds is consistent with the design goal of allocating inference cost by problem difficulty.

\textbf{Escalation targets instances that are difficult for direct inference.}
To test whether escalation merely reflects random fluctuation, we contrast the 26 escalated instances with the direct-inference baseline: the baseline's error rate on these instances is 76.9\%, substantially higher than its 50.63\% over all samples, or about 1.5 times as high. This indicates that the instances triggering extra rounds are predominantly those that remain difficult for direct inference without any optimization. The placement of depth aligns with instance difficulty as measured by direct inference rather than with random variation; escalation does not guarantee solving these hardest instances, but the extra interaction is invested where it is needed most.

\textbf{The single failure case.}
Across the evaluation trajectories, exactly one failed to complete the interaction, a rate of 0.05\%: on one ScienceQA sample, the agent emitted a closing tag with one letter missing, the parser therefore recognized no valid steering prompt, the target model was never called, and the sample was counted as a failure. Such errors occur at the level of tag spelling in the output rather than at the level of reasoning; they are deterministically detectable by the parser and can be prevented by constrained decoding or caught by tag validation.

\section{Training Compute Cost}
\label{app:cost}

Training runs on the server with eight NVIDIA H100 80GB GPUs for 90 steps over a single epoch, with a batch size of 128 and four sampled trajectories per input, totaling 46,080 training trajectories; over the course of training, the agent issues approximately 100.5K calls to the target MLLM, which processes approximately 0.17B tokens. The end-to-end wall-clock time is approximately 3.58 days, about 687 GPU-hours for one end-to-end training run on the eight-GPU server, with four GPUs training the agent and four serving the frozen target model. This cost is a one-time investment: the trained policy requires no further training for inference or for cross-model transfer.

\section{Answer Reward Robustness}
\label{app:reward-robust}

The answer reward measures the match between a prediction and the ground truth by token-level F1, which could in principle push the policy toward shortening answers to gain precision. To test this risk, this section compares the answer-length distributions of the ground truth, our method, and the direct-inference baseline in Table~\ref{tab:length}.

\begin{table}[!htb]
\centering
\caption{Answer length distributions and overall accuracy. Mean, P90, and Max are the average, the 90th percentile, and the maximum of the answer length in words; the baseline EM and F1 are sample-weighted averages over the fifteen datasets.}
\label{tab:length}
\vspace{-0.6em}
{\fontsize{8}{9}\selectfont
\begin{tblr}{colspec={Q[l,2.4cm]Q[c,0.8cm]Q[c,0.7cm]Q[c,0.7cm]Q[c,0.8cm]Q[c,0.8cm]},
  colsep=3pt,rowsep=1.5pt,stretch=1.0,
  row{1}={bg=white, font=\bfseries\fontsize{8}{9}\selectfont},
  row{2-3}={font=\fontsize{8}{9}\selectfont},
  row{4}={font=\fontsize{8}{9}\selectfont},
  row{3}={font=\bfseries\fontsize{8}{9}\selectfont, bg=grpC},
  hline{1,Z}={1.0pt},
  hline{2}={0.6pt}}
\textbf{Answer} & \textbf{Mean} & \textbf{P90} & \textbf{Max} & \textbf{EM} & \textbf{F1} \\
Ground truth & 3.1 & 7 & 40 & -- & -- \\
Ours & 4.1 & 10 & 70 & 59.48 & 71.77 \\
Baseline & 4.9 & 11 & 116 & 49.37 & 63.85 \\
\end{tblr}}
\end{table}

\textbf{Length is calibrated to the reference, not collapsed.}
Across all three statistics, the answer length of our method is the closest to the ground-truth distribution: moderately longer than the reference, shorter than the baseline, and with the longest answer far below the verbose tail of the baseline. If the reward had induced brevity gaming, lengths systematically below the ground truth would be expected. The observation points in the opposite direction: the policy learns an answering length aligned with the reference style rather than degenerate minimal outputs.

\textbf{More concise and more accurate.}
Relative to direct inference, our method answers more concisely while scoring higher on both accuracy metrics. The gains are consistent with improved correctness rather than with inflating token overlap through longer answers or trimming answers to harvest precision. In practice, we observe no sign of a systematic length bias induced by the answer reward.

\section{Limitations}
\label{app:limitations}
Although MMDynOpt-Agent improves the reasoning performance of target MLLMs through multi-turn dynamic optimization, it still has several applicability boundaries. \textbf{First}, because it relies on multi-turn interaction to progressively refine the reasoning process, it is better suited to scenarios that prioritize reasoning quality and permit limited interaction budget; under strict single-turn, ultra-low-latency, or ultra-low-cost settings, its optimization space is constrained. \textbf{Second}, MMDynOpt-Agent is an external reasoning optimization mechanism that guides the target MLLM by converting visual cues and question semantics into more effective reasoning conditions, rather than directly improving the model’s parameters or capabilities. As a result, its gains remain bounded by the target model’s visual perception, knowledge, instruction following, and reasoning abilities. \textbf{Finally}, the framework depends on a unified interaction interface and stable multi-turn behavior. Differences across target MLLMs in instruction sensitivity, context retention, and interaction consistency may therefore affect the effectiveness and transferability of the dynamic optimization strategy.

\section{Future Work}
\label{app:future}
Future work can proceed along several directions. \textbf{First}, since multi-turn dynamic optimization still incurs interaction overhead, future studies may investigate more efficient adaptive mechanisms, including finer-grained early stopping, budget-aware decision policies, and difficulty-adaptive control of interaction rounds, to further reduce latency and cost while preserving reasoning quality. \textbf{Second}, because this framework operates as an external reasoning optimization mechanism and remains constrained by the capabilities of the target MLLM, future research may integrate dynamic optimization with stronger visual encoding, retrieval augmentation, tool use, or domain knowledge injection, so that the framework can not only optimize reasoning conditions but also mitigate limitations in perception, knowledge, and complex reasoning. \textbf{Finally}, to address heterogeneity across target MLLMs in interface design, context retention, and multi-turn consistency, future work may explore more model-agnostic dynamic optimization strategies and interface adaptation mechanisms, thereby improving framework stability, compatibility, and transferability across models and tasks.

\end{document}